\documentclass[useAMS,usenatbib]{mnras}
\usepackage{graphicx}
\usepackage{amsmath, amssymb}
\usepackage[T1]{fontenc}
\usepackage[usenames]{color}
\usepackage[dvipsnames]{xcolor}
\usepackage{bm}
\usepackage[capitalize]{cleveref}
\usepackage{physics}
\usepackage{acro}
\usepackage{caption}
\usepackage{subcaption}
\usepackage{stfloats}
\usepackage{xspace}
\usepackage{hyperref}
\usepackage{booktabs}
\usepackage{tabularx}
\usepackage{multirow}
\usepackage{acro}
\usepackage{lmodern}
\usepackage[normalem]{ulem}
\usepackage{orcidlink}

\newcommand{\zcmb}{z_{\rm obs}}
\newcommand{\zpred}{z_{\rm pred}}

\newcommand{\zCMB}{z_{\rm CMB}}
\newcommand{\zcosmo}{z_{\rm cosmo}}
\newcommand{\zpec}{z_{\rm pec}}

\newcommand{\vpec}{V_{\rm pec}}

\newcommand{\mutrue}{\mu_{\rm true}}

\newcommand{\muTFR}{\mu_{\rm TFR}}

\newcommand{\mobs}{m_{\rm obs}}
\newcommand{\mtrue}{m_{\rm true}}

\newcommand{\etaobs}{\eta_{\rm obs}}

\newcommand{\etatrue}{\eta_{\rm true}}

\newcommand{\sigmaTFR}{\sigma_{\rm TFR}}
\newcommand{\aTFR}{a_{\rm TFR}}
\newcommand{\bTFR}{b_{\rm TFR}}
\newcommand{\cTFR}{c_{\rm TFR}}

\newcommand{\Om}{\Omega_{\rm m}}
\newcommand{\RLG}{R_{\rm offset}}
\newcommand{\Rvoid}{\tilde{r}_{\rm void}}

\DeclareAcronym{LG}{short = LG, long = Local Group}
\DeclareAcronym{TFR}{short = TFR, long  = Tully--Fisher relation}
\DeclareAcronym{CF4}{short = CF4, long  = CosmicFlows-4}
\DeclareAcronym{SDSS}{short = SDSS, long = Sloan Digital Sky Survey}
\DeclareAcronym{WISE}{short = WISE, long = Wide-field Infrared Survey Explorer}
\DeclareAcronym{CMB}{short = CMB, long  = cosmic microwave background}
\DeclareAcronym{LCDM}{short = $\Lambda$CDM, long  = $\Lambda$-cold dark matter}
\DeclareAcronym{LOS}{short = LOS, long  = line-of-sight}
\DeclareAcronym{HMC}{short = HMC, long  = Hamiltonian Monte Carlo}

\defcitealias{Haslbauer_2020}{HBK20}
\defcitealias{Watkins_2023}{W23}
\defcitealias{Mazurenko_2024}{M24}
\defcitealias{Jia_2023}{JHW23}
\defcitealias{Jia_2024}{JHW24}
\defcitealias{VF_olympics}{S25}

\title[Testing the local void with distance tracers]{Testing the local supervoid solution to the Hubble tension with direct distance tracers}
\author[Stiskalek et al.]{Richard Stiskalek,$^{1,2}$\thanks{\href{mailto:richard.stiskalek@physics.ox.ac.uk}{richard.stiskalek@physics.ox.ac.uk}}\orcidlink{0000-0002-0986-314X}
Harry~Desmond$^{3}$\orcidlink{0000-0003-0685-9791}
and Indranil Banik$^3$\orcidlink{0000-0002-4123-7325}\\
$^{1}$Department of Astrophysics, University of Oxford, Denys Wilkinson Building, Keble Road, Oxford, OX1 3RH, UK\\
$^{2}$Center for Computational Astrophysics, Flatiron Institute, 162 5\textsuperscript{th} Ave, New York, NY 10010, USA\\
$^{3}$Institute of Cosmology \& Gravitation, University of Portsmouth, Dennis Sciama Building, Burnaby Road, Portsmouth PO1 3FX, UK}

\date{Accepted XXX. Received YYY; in original form ZZZ}
\pubyear{\the\year{}}

\begin{document}\label{firstpage}
\pagerange{\pageref{firstpage}--\pageref{lastpage}}
\maketitle

\begin{abstract}
Several observational studies suggest that the local few hundred Mpc around the Local Group is significantly underdense based on source number counts in redshift space across much of the electromagnetic spectrum, particularly in near-infrared galaxy counts. This ``Keenan--Barger--Cowie (KBC) void'', ``Local Hole'', or ``local supervoid'' would have significant ramifications for the Hubble tension by generating outflows that masquerade as an enhanced local expansion rate. We evaluate models for the KBC void capable of resolving the Hubble tension with a background \emph{Planck} cosmology. We fit these models to direct distances from the Tully--Fisher catalogue of the CosmicFlows-4 compilation using a field-level forward model. Depending on the adopted void density profile, we find the derived velocity fields prefer a void size $\la70$~Mpc, which is $\la10$ per cent of the fiducial size found by Haslbauer et al.~based on the KBC luminosity density data. The predicted local Hubble constant is $72.1^{+0.9}_{-0.8}$, $70.4^{+0.4}_{-0.4}$, or $70.2^{+0.5}_{-0.4}$~km/s/Mpc for an initial underdensity profile that is exponential, Gaussian, or Maxwell-Boltzmann, respectively. The latter two ameliorate the Hubble tension to within $3\sigma$ of the 4-anchor distance ladder approach of Breuval et al., which gives $73.2 \pm 0.9$~km/s/Mpc. The exponential profile achieves consistency with this measurement at just over $1\sigma$, but it is disfavoured by the Bayesian evidence. The preferred models produce bulk flow curves that disagree with recent estimates from CosmicFlows-4, despite the void models being flexible enough to match such estimates.
\end{abstract}



\section{Introduction}\label{sec:intro}

One of the biggest current puzzles in cosmology is the Hubble tension, a statistically significant discrepancy between the rate at which redshift increases with distance as inferred from the local distance ladder versus the predicted rate assuming the \ac{LCDM} model with parameters calibrated using the \ac{CMB} anisotropies. In a homogeneously expanding universe, the redshift gradient $z' \equiv \dd z/ \dd r$ with respect to distance $r$ arises purely from cosmic expansion, implying that $cz' = \dot{a}$, where $c$ is the speed of light, $a$ is the cosmic scale factor, and an overdot denotes a time derivative \citep[for a more detailed derivation, see the introduction to][]{Mazurenko_2025}. The expansion rate is quantified by the Hubble parameter $H \equiv \dot{a}/a$, leading to $cz' = H_0$ at the present day when $a \equiv 1$ (a subscript 0 denotes evaluation at $z=0$). Since the overall expansion history $a(t)$ as a function of time $t$ since the Big Bang can be predicted in \ac{LCDM}, it is then possible to predict the local $cz'$ using a calibration at early times based on the \ac{CMB} anisotropies. The most precise such results are derived from the \textit{Planck} mission~\citep{Planck_2006}---${H_0^{\mathrm{Planck}} = 67.4 \pm 0.5}$~km/s/Mpc~\citep{Planck_2020, Tristram_2024}---with similar values obtained from ground-based telescopes~\citep{Calabrese_2025, Camphuis_2025, Ge_2025}. However, $cz'$ measured from the local distance ladder is persistently higher~\citep[for a review, see][]{Valentino_2020_discordance}. The most precise local measurement is ${H_0^{\rm local} = 73.17\pm0.86}$~km/s/Mpc~\citep{Breuval_2024} from the SH0ES programme~\citep{Riess_2021}, derived by calibrating the absolute magnitude of type Ia supernovae (SNe~Ia) with the Cepheid period--luminosity relation~\citep{Leavitt_1912} in four anchor galaxies with trigonometric distances. This is in $6\sigma$ tension with the CMB value, but agrees with other methods of building the local distance ladder without Cepheids, SNe~Ia, or both~\citep{Pesce_2020, Scolnic_2023, Bhardwaj_2025, Jensen_2025, Vogl_2025}.

It has been suggested that the Milky Way resides in a large local underdensity or void, which would violate the relation $cz' = \dot{a}$ by providing an additional contribution to $z'$ from peculiar velocity and gravitational redshift due to our location on a potential hill~\citep{Keenan_2016, Shanks_2019a, Shanks_2019b, Ding_2020, Camarena_2022, Martin_2023, Cai_2025}. A local void was in fact proposed long before the Hubble tension based on galaxy number counts across the whole electromagnetic spectrum. The claims for the largest and deepest void come from galaxy number counts in the near-infrared~\citep*{Keenan_2013}: the ``KBC void''. This led~\citealt*{Haslbauer_2020} (hereafter~\citetalias{Haslbauer_2020}) to construct a semi-analytic model relating the KBC void to the Hubble tension. Accounting for excess redshift induced by the void,~\citetalias{Haslbauer_2020} argued that the KBC void is roughly 20~per cent underdense out to 300~Mpc and causes the local $cz'$ to exceed $\dot{a}$ by $11 \pm 2$ per cent (see their Equation~5). This could fully solve the Hubble tension while retaining the \emph{Planck} cosmology at the background level, i.e., the present $\dot{a} = H_0^{\mathrm{Planck}}$~\citep[see Figure~3 of][]{Valentino_2025}.

An important result of~\citetalias{Haslbauer_2020} was that the observed size and depth of the KBC void are inconsistent with \ac{LCDM} expectations at $6\sigma$, regardless of the absolute distance scale. This is because the \ac{LCDM} power spectrum cosmic variance on 300~Mpc scales is only 4.8 per cent in redshift space, but observations across 90 per cent of the sky and most ($57-75$~per cent) of the luminosity function suggest an underdensity of $46 \pm 6$ per cent~\citep[see the light blue point on Figure~11 of][]{Keenan_2013}. The lack of sufficient structure in $\Lambda$CDM on the relevant scales prevents it from solving the Hubble tension through cosmic variance~\citep{Wu_2017}, which is expected to be only $\approx 0.9$~km/s/Mpc in the local $cz'$ given the scales over which this is typically measured~\citep{Camarena_2018}. As a result, a local void in the \ac{LCDM} framework cannot solve the Hubble tension or explain the KBC void. However, if structure formation proceeded faster than expected in \ac{LCDM} on scales $\ga 100$~Mpc, cosmic variance might be enhanced sufficiently to solve both problems. The models of~\citetalias{Haslbauer_2020} were therefore based on Modified Newtonian dynamics~\citep[MOND;][]{Milgrom_1983, Famaey_McGaugh_2012, Banik_Zhao_2022} as a convenient way to boost the gravitational field from a given density distribution; this is to be considered simply as a proxy for a theory with stronger gravity than \ac{LCDM}.~\citetalias{Haslbauer_2020} considered initial underdensities at $z = 9$ ($a = 0.1$) conforming to exponential, Gaussian, and Maxwell-Boltzmann profiles, finding in each case the parameters which best fit the local redshift gradient and curvature from supernovae, the KBC void density profile inferred from galaxy number counts, strong lensing time delays, and the peculiar velocity of the \ac{LG}, which needs to be plausible given the predicted void velocity field. The models provided a satisfactory explanation for the observables, achieving an overall Gaussian-equivalent tension of $2.5\sigma$, $2.8\sigma$, and $2.9\sigma$ for the Maxwell-Boltzmann, Gaussian, and exponential profile, respectively.

The scope of the void model was extended in~\citealt{Mazurenko_2024} (hereafter~\citetalias{Mazurenko_2024}) by fitting it to the bulk flow, a measure of the average velocity of galaxies within spheres of varying radii centred on our location~\citep{Watkins_2009, Feldman_2010, Nusser_2011, Hoffman_2015, Watkins_2015, Nusser_2016, Scrimgeour_2016, Feix_2017, Hellwing_2018, Peery_2018}. This was motivated by an inference of the bulk flow by~\citealt{Watkins_2023}~\citepalias[hereafter][]{Watkins_2023}, which indicated tension with the \ac{LCDM} expectation at the $5\sigma$ level. The bulk flow was measured using the \ac{CF4} compilation of redshifts and redshift-independent distances~\citep{Tully_2023}, finding increasing tension out to $250 ~ \mathrm{Mpc} / h$. The bulk flow magnitude claimed by~\citetalias{Watkins_2023} was later corroborated by~\citet{Whitford_2023}, though they were only able to probe out to $173 \, \mathrm{Mpc} / h$ due to a more conservative methodology. They also argued that the uncertainties in the bulk flow claimed by~\citetalias{Watkins_2023} were underestimated by a factor of $\approx 3.5$.~\citetalias{Mazurenko_2024} showed that the void model calibrated in~\citetalias{Haslbauer_2020} can explain the anomalous CF4 bulk flow curve reported by~\citetalias{Watkins_2023}, but this is based on a misunderstanding in~\citetalias{Mazurenko_2024} which alters their conclusions regarding the preferred void properties and our location in the void.\footnote{This is acknowledged and rectified by all those authors in a subsequent study (Mazurenko et al., submitted).} It is therefore important to revisit the issue of whether the local velocity field is consistent with the void model.

While our focus is at low $z$, an important consequence of a local void model is that we expect a return to the \textit{Planck} cosmology at high $z$. The rate of convergence was explored in more detail by~\citet{Mazurenko_2025} in terms of $H_0 \left( z \right)$, the value of $H_0$ that would be inferred using data in a narrow redshift range centred on $z$. They found good agreement with the observationally inferred $H_0 \left( z \right)$~\citep{Jia_2023, Jia_2025}. The expectation that high redshift datasets are consistent with the \textit{Planck} cosmology can be assessed using cosmic chronometers (CCs), which have very little sensitivity to $z \la 0.3$ because the CC technique involves finding $\dot{z}$ by comparing the differential age between galaxy populations at two redshifts~\citep[][and references therein]{Moresco_2024}. Results using CCs combined with other uncalibrated datasets lead to $H_0$ values very close to $H_0^{\mathrm{Planck}}$ and below the local $cz'$~\citep{Cogato_2024, Guo_2025}. This is also the case with strong lensing time delays of supernova Refsdal~\citep{Kelly_2023, Grillo_2024}. Moreover, the ages of the oldest Galactic stars are much more in line with the low $H_0^{\mathrm{Planck}}$~\citep{Cimatti_2023, Valcin_2025, Xiang_2025}. For a summary of these and other constraints on $H_0$ and the matter density parameter $\Om$, we refer the reader to \citet{Banik_2025_no_CMB}.

The local void scenario was challenged by~\citet{Kenworthy_2019} using the distance-redshift relation of SNe~Ia. However those authors assumed that a local void would have essentially no effect on redshifts at $z > 0.1$ (see their Figure~1), which is not the case in the models of~\citetalias{Haslbauer_2020}~\citep[see Figure~3 of][]{Mazurenko_2025}. It may also be relevant that SNe~Ia are not standard candles, but merely standardisable~\citep{Phillips_1993, Tripp_1998, Brout_2022}. The standardisation procedure can introduce dependence on the cosmological model~\citep{Lane_2025}, making it important to jointly infer the cosmological parameters with the calibration parameters~\citep{Seifert_2025}. This may cast doubt on studies that argue against a local void using SNe~Ia while calibrating their Tripp parameters using a void-free model~\citep{Camarena_2022, Castello_2022}. These issues can be avoided by using instead baryon acoustic oscillation (BAO) measurements over the last twenty years, which show a preference for a local void~\citep{Banik_2025_BAO} or other late-time adjustment to the \textit{Planck} cosmology such as dynamical dark energy~\citep{DESI_2025}, though the latter approach worsens the Hubble tension \citep{Mirpoorian_2025}.

This work focuses on low $z$ and extends the analysis of~\citetalias{Mazurenko_2024} by analysing velocities at the field-level (i.e. galaxy-by-galaxy), rather than relying solely on the reported bulk flow curve. The advantage of this is twofold. First, it lets us extract more information from the data without compressing it into a summary statistic. Second, it lets us avoid assumptions about the accuracy of the bulk flow produced by~\citetalias{Watkins_2023}, including the accuracy of their error model~\citep[questioned by][]{Whitford_2023}. Indeed, a field-level inference allows us to check the bulk flow estimate of~\citetalias{Watkins_2023} by investigating consistency between constraints on the void velocity field inferred by fitting to the~\citetalias{Watkins_2023} bulk flow versus fitting the galaxies individually and deriving the posterior predicted bulk flow from the latter. In particular, we use the \acl{TFR}~\citep[\ac{TFR};][]{Tully_1977} sample from the \ac{CF4} survey~\citep{Kourkchi_2020B, Kourkchi_2020A, Tully_2023}, deriving constraints on the void profile, size, the observer offset from the void centre, and the velocity with which the void moves with respect to the rest of the universe.

In~\cref{sec:data}, we explain the data we use and the quality cuts we apply. Our methods are described in~\cref{sec:methods}. We then present our results in~\cref{sec:results} and discuss them in~\cref{sec:discussion}. We conclude in~\cref{sec:conclusions}. All logarithms in this work are base 10. We use the notation $\mathcal{N}(x; \mu, \sigma)$ to denote the normal distribution with mean $\mu$ and standard deviation $\sigma$ evaluated at $x$.

\section{Observational Data}\label{sec:data}

The \ac{TFR} is an empirical relation between the width $W$ of a spectral line in a spiral galaxy (typically the 21~cm H\textsc{I} line of neutral hydrogen) as a measure of its rotation velocity, and absolute magnitude $M$ as a measure of its luminosity~\citep{Tully_1977}. We write the relation as
\begin{equation}
    M =
    \begin{cases}
        \aTFR + \bTFR \eta + \cTFR \eta^2\quad~&\mathrm{if}~\eta > 0,\\
        \aTFR + \bTFR \eta\quad~&\mathrm{otherwise},
    \end{cases}
\label{eq:TFR_absmag}
\end{equation}
where $\aTFR$, $\bTFR$, and $\cTFR$ are the zero-point, slope, and curvature, respectively. We reparametrise the linewidth by introducing a parameter $\eta$ such that
\begin{equation}
    \eta \equiv \log \frac{W}{\mathrm{km/\mathrm{s}}} - 2.5.
    \label{eq:eta_definition}
\end{equation}
From now on, we refer to $\eta$ as the linewidth. In the subsequent analysis, we will jointly infer the \ac{TFR} calibration parameters along with the intrinsic scatter $\sigmaTFR$. Curvature is included at $\eta > 0$, where the relation is empirically found to deviate from linearity.

\begin{figure}
    \centering
    \includegraphics[width=\columnwidth]{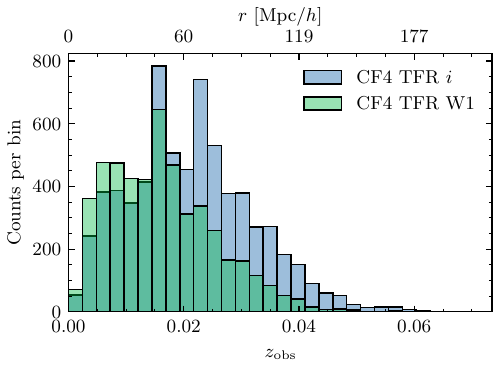}
    \caption{The distribution of galaxy redshifts in the \ac{CMB} frame ($\zcmb$) for the \ac{CF4} \ac{TFR} sample. The upper $x$-axis shows $\zcmb$ converted to comoving distance assuming no peculiar velocity. The plot shows separately the set of galaxies with $i$-band \ac{SDSS} photometry and $W1$-band \ac{WISE} photometry.}
    \label{fig:CF4_redshift_sky_dist}
\end{figure}

We use the \ac{CF4} \ac{TFR} survey, which contains $9,792$ galaxies at $z < 0.065$, with $5,027$ in the \ac{SDSS} $i$-band subsample and $3,278$ in the \ac{WISE} $W1$ subsample~\citep{Kourkchi_2020B, Kourkchi_2020A, Tully_2023}. We show the redshift distribution of the sample in~\cref{fig:CF4_redshift_sky_dist}. Our selection includes galaxies with $\eta > -0.3$, Galactic latitude $|b| > 7.5^\circ$, and a photometric quality flag of $5$ (the best quality). The cut in $\eta$ eliminates dwarf and low-mass galaxies, which are difficult to detect and may follow a different \ac{TFR} or have higher scatter. In addition, their peculiar velocities are more strongly affected by non-linear motions (e.g.~as satellites) and thus do not provide a reliable probe of the large-scale velocity field. The cut in $b$ is to mask out sources close to the Galactic Zone of Avoidance. Lastly, we exclude 168 galaxies identified as \ac{TFR} outliers by~\citet{Boubel_2024}. We use the \ac{SDSS} photometry when available and \ac{WISE} otherwise. We assign distinct \ac{TFR} parameters for each photometric system, since differences in photometry or calibration result in distinct TFRs. Following the selection, the joint sample contains $5,025$ and $2,094$ galaxies with \ac{SDSS} and \ac{WISE} photometry, respectively.

Furthermore, we adopt the Cepheid-based absolute calibration from the SH0ES catalogue~\citep{Riess_2022_comprehensive}. Out of the 40 galaxies with Cepheid-derived distances, we match 18 of them to the \ac{CF4} \ac{TFR} catalogue, of which only 12 remain in the \ac{CF4} sample after applying the selection criteria described above. The consistency of the SH0ES Cepheid calibration has been recently investigated by~\citet{Najera_2025}, who found no evidence for unaccounted-for systematics. However, we emphasize that we do not infer the value of the Hubble constant directly --- this is implicitly fixed to the \emph{Planck} value, which defines the background cosmology used to construct the void model. Instead, the absolute calibration is used to break the degeneracy between a radial outflow induced by the void and a global zero-point offset in the Tully--Fisher normalization.


\section{Methodology}\label{sec:methods}

Here we introduce our methodology to set constraints on the local void from the peculiar velocity data. The model of void growth is from~\citetalias{Haslbauer_2020}, with fixed initial density contrast and variable parameters including the void’s external velocity, void size, and observer position (Section~\ref{sec:void_model}). We first derive constraints based on the bulk flow curve, incorporating an important correction for projection effects that was previously neglected (Section~\ref{sec:bf_model}). We then describe the main method of this paper: a full galaxy-by-galaxy Tully--Fisher flow model to jointly infer void and calibration parameters (Section~\ref{sec:flow_model}). An additional constraint from the CMB quadrupole is described in Appendix~\ref{sec:CMB_quadrupole}: the off-centre lensing contribution to large-scale anisotropies must remain subdominant, imposing an upper bound on the induced quadrupole amplitude ($a_{20} < 10^{-5}$), which significantly restricts the allowed void parameter space. These constraints are particularly relevant for the fiducial void size considered in~\citetalias{Haslbauer_2020}, implying that the observer must lie within $\approx 90~\mathrm{Mpc}$ of the void centre for all three profiles. However, the CMB quadrupole does not impose significant constraints on the smaller voids preferred by our field-level analysis.

\subsection{Void model}\label{sec:void_model}

The relation between the density contrast of a local void and the outflow it induces depends on the void's growth history. Standard equations cannot consistently describe this growth in the case of the KBC void because its properties are in $6.0\sigma$ tension with \ac{LCDM} expectations, an issue that is observationally unrelated to the Hubble tension~\citepalias[Section~2.2 of][]{Haslbauer_2020}. In \ac{LCDM}, accurate measurements of the CMB anisotropies provide a precise calibration of the early Universe density contrasts on the relevant scale of 300 comoving Mpc (cMpc), which in turn fixes the expected density fluctuations on that scale today. Since this expectation is contradicted by the observed density contrast of the KBC void, those authors were motivated to adopt a semi-analytic model with enhanced structure growth based on the MOND framework \citep{Milgrom_1983}. This was used to evolve an initial density perturbation from $a = 0.1$ to the present epoch ~\citepalias[see Section~3.2 of][]{Haslbauer_2020}.

Those authors considered three initial underdensity profiles: exponential, Gaussian, and Maxwell-Boltzmann. This was to allow exploration of how the density profile might impact the results. In the exponential and Gaussian profiles, the void is deepest at its centre, which is more in line with the observed density profiles of simulated and observed voids on smaller scales \citep{Nadathur_2015, Curtis_2025}. The Maxwell-Boltzmann profile has a density profile that climbs back to the cosmic mean value at the void centre, placing the deepest part of the void somewhat off-centre. This unusual arrangement appears not to work well with the bulk flow curve \citepalias{Mazurenko_2024}, but is nonetheless considered here for completeness. It was originally considered in \citetalias{Haslbauer_2020} based on the observations of \citet{Karachentsev_2018}, which suggest that the density within 11~Mpc is larger than the density within 40~Mpc, implying that the density first decreases away from us before then rising back up. However, the number of galaxies within 11~Mpc is quite small, making it unclear how reliable these results are. Moreover, there is an obvious anthropic selection effect in that we need to reside within a galaxy, implying that the average density on sufficiently small scales must rise, even if we are within a supervoid.

We model the growth history of the void using the same approach and fix the initial void density contrast to the best-fit values in Tables~4 and C1 of~\citetalias{Haslbauer_2020}. However, we treat the systemic velocity of the void relative to the rest of the Universe ($\bm{V}_{\mathrm{ext}}$), the void size relative to its fiducial value ($\tilde{r}_{\rm void}$), and our offset from the void centre ($\RLG$) as free parameters (\citetalias{Haslbauer_2020} were unable to constrain the latter). The models considered are spherically symmetric outflow models, but the introduction of $\bm{V}_{\mathrm{ext}}$ reduces this symmetry to axisymmetry. To transform into the CMB frame, we apply an additional Galilean transformation. To avoid added complexity, we restrict our vantage point to positions along the void symmetry axis~\citepalias[similarly to][]{Mazurenko_2024}.

The best-fitting initial void size from~\citetalias{Haslbauer_2020} is 1030~cMpc for both the exponential and Gaussian profiles, but only 228.2~cMpc for the Maxwell-Boltzmann profile (see their Tables~4 and C1). To relax this assumption, we scale the void size relative to these values by $\tilde{r}_{\rm void}$, which we treat as a free parameter. Additionally, to mimic the effect of varying the void depth, we scale the void-predicted peculiar velocities by the factor $\beta$, which we sample.

\subsection{Updated constraints from the bulk flow curve}\label{sec:bf_model}

An important constraint on the local void model is the predicted $\bm{V}_{\rm pec}$ of each observed galaxy with a redshift-independent distance. One way to describe the velocity field of the local Universe is to consider all galaxies within a sphere of radius $R$ centred on our location and compute their average velocity. This is the bulk flow $\bm{V}_b$. Taking the average within a spherical region would in ideal circumstances completely eliminate dependence on the assumed $H_0$, whose value is highly controversial yet necessary to calculate an individual galaxy's peculiar velocity.  Observationally, $\bm{V}_b$ must be calculated using only \ac{LOS} peculiar velocities $V_{\rm pec}$. These can be treated as vectors pointing along the \ac{LOS}, such that $\bm{V}_{\rm pec}^{\rm LOS} \equiv V_{\rm pec} ~\hat{\bm{r}}$, where $\hat{\bm{r}}$ is the unit vector towards the galaxy.

Recently,~\citetalias{Watkins_2023} measured the bulk flow curve $\bm{V}_b \left( R \right)$ out to $250 \, \mathrm{Mpc}/h$. This motivated~\citetalias{Mazurenko_2024} to compare their measurements to the predictions of the local void model.~\citetalias{Mazurenko_2024} discuss this procedure in their Section~2.3, with their Equation~4 stating that
\begin{eqnarray}
    \bm{V}_b = \frac{\sum_i w_i \bm{V}_{\mathrm{pec}}^{\mathrm{LOS}}}{\sum_i w_i},
    \label{eq:Mazurenko_bulk_flow}
\end{eqnarray}
where $w_i$ is the relative statistical weight assigned to the $i$\textsuperscript{th} galaxy, which in the model corresponds to the volume of each cell. Since the void models considered by~\citetalias{Mazurenko_2024} were axisymmetric, it was clear that $\bm{V}_b$ would point along the symmetry axis, so only this component of $\bm{V}_{\mathrm{pec}}^{\mathrm{LOS}}$  had to be considered. The consideration of only one axis direction is motivated by all the observationally reported bulk flows from~\citetalias{Watkins_2023} being aligned with each other to within a few degrees. This common direction can be taken as the symmetry axis of the void model given the lack of any other constraints on how it should be oriented.

\citetalias{Mazurenko_2024} pointed out that due to the need to project peculiar velocities onto the \ac{LOS} and then onto the symmetry axis, $\bm{V}_b$ calculated in this way will be only $\bm{V}_{\mathrm{pec}}/3$ if all galaxies in the considered volume have the same $\bm{V}_{\mathrm{pec}}$. However, this approach is not how the bulk flow is defined in the literature. Indeed, Equation~2 of~\citet{Nusser_2016} recovers Equation~\ref{eq:Mazurenko_bulk_flow}, but with an extra factor of three. This is designed to mitigate the above-mentioned projection issues, so the estimated $\bm{V}_b = \bm{V}_{\mathrm{pec}}$ if $\bm{V}_{\mathrm{pec}}$ is the same for all galaxies, making the results much simpler to interpret. To illustrate the origin of the factor of three, consider the relation between the constant bulk velocity $\bm{V}_b$ and its \ac{LOS} projection for a top-hat filter, where the bulk flow is defined as the volume average of the \ac{LOS} velocity treated as a vector pointing along the \ac{LOS}.
\begin{equation}
    \bm{V}_b = 3 \times\frac{1}{V} \int \mathrm{d}V \,\left(\bm{V}_b \cdot \hat{\bm{r}}\right)\hat{\bm{r}},
\end{equation}
where $V$ is the enclosed volume and $\dd V$ is the volume element. This identity holds only when the factor of three is included, which is not the case in Equation~\ref{eq:Mazurenko_bulk_flow}. As a result, \emph{all the bulk flow predictions in~\citetalias{Mazurenko_2024} must be tripled} (as acknowledged by those authors in Mazurenko et al., submitted). The corrected formula is therefore
\begin{eqnarray}
    \bm{V}_b = \frac{3\sum_i w_i \bm{V}_{\mathrm{pec}}^{\mathrm{LOS}}}{\sum_i w_i}.
    \label{eq:bulk_flow}
\end{eqnarray}

\citetalias{Mazurenko_2024} did not vary the void model parameters and only considered a very limited number of possible vantage points. Given the bulk flow curves were found to be in good agreement with the measurements of~\citetalias{Watkins_2023}, it is clear that tripling the model predictions to account for the above mistake would lead to a very poor agreement. To check if the local void scenario might still be plausible, in this work we explore a much wider range of model parameters, varying not only our vantage point but also the void external velocity $\bm{V}_{\mathrm{ext}}$ (Section~\ref{sec:void_model}). In each case, we compute the $\chi^2$ as the minimum of $\chi_+^2$ and $\chi_-^2$, where
\begin{eqnarray}
    \chi_\pm^2 = \sum_i \left( \frac{V_{b,i}^{\mathrm{obs}} \mp V_{b,i}^{\mathrm{pred}}}{\sigma \left( V_{b,i}^{\mathrm{obs}} \right)} \right)^2 \, ,
    \label{eq:chi_sq_bf}
\end{eqnarray}
where the sum runs over the radial bins indexed by $i$. The model predictions and observations are distinguished using ``pred'' and ``obs'' superscripts, respectively, with the latter having an uncertainty of $\sigma \left( V_b^{\mathrm{obs}} \right)$. The two possible choices of sign arise because unlike the observed bulk flow curve where the directions are all very nearly parallel, the predicted bulk flow curve could reverse direction~\citepalias[this occurs for the outer vantage points in Figure~1 of][]{Mazurenko_2024}. Since there is no \emph{a priori} way to know if $\bm{V}_{\mathrm{ext}}$ is parallel to the observed $\bm{V}_b$ found by~\citetalias{Watkins_2023} or in the opposite direction, we can orient the void model as we wish. We therefore compute $\chi_+^2$ and $\chi_-^2$ using both the predicted bulk flow curve $\bm{V}_b^{\mathrm{pred}} \left( R \right)$ and its negative $-\bm{V}_b^{\mathrm{pred}} \left( R \right)$, respectively, and then find $\min \left( \chi_+^2, \chi_-^2 \right)$. We assume for simplicity that the measurements of~\citetalias{Watkins_2023} are all independent and yield the same direction. However, even if the bulk flow direction is constant, the independence assumption is clearly violated, as the bulk flow represents a cumulative measurement of the mean velocity within concentric spheres. This will introduce some systematic error into this part of the analysis. We also take advantage of the two extra points at low radii shown in Figure~1 of~\citetalias{Mazurenko_2024}. It is clear that if there are model parameters which give an acceptable $\chi^2$ in this sense, the parameters must be quite different to those considered by~\citetalias{Mazurenko_2024}, so a successful fit can no longer be considered a successful \emph{a priori} prediction of the model.

\subsection{Object-by-object velocity field model}
\label{sec:flow_model}

To extract more of the information contained in the direct-distance catalogues, we perform a galaxy-by-galaxy fit without first compressing the data into a bulk flow. We simultaneously calibrate the \ac{TFR} and infer the parameters of the void model, which predicts both the density and the radial velocity. In the \ac{CMB} frame, the total source redshift $\zCMB$ is given by
\begin{equation}
    1 + \zCMB = \left(1 + \zcosmo\right)\left(1 + \zpec\right),
    \label{eq:redshift_addition}
\end{equation}
where $\zcosmo = a^{-1} - 1$ is the redshift due to cosmic expansion and $\zpec = \vpec / c$ is the redshift due to the radial peculiar velocity $\vpec$, also measured in the \ac{CMB} frame (see e.g. Section 2.3 of~\citealt{Davis_2019}). We assume a flat \ac{LCDM} universe dominated by non-relativistic matter and dark energy to convert between the cosmological redshift and the comoving distance~\citep[as described by, e.g.,][]{Hogg_1999},
\begin{equation}
    r(\zcosmo) = \frac{c}{H_{\rm 0}} \int_{0}^{\zcosmo} \frac{\dd z^\prime}{\sqrt{\Om (1 + z^\prime)^3 + (1 - \Om)}},
    \label{eq:redshift_to_distance}
\end{equation}
where $\Om$ is the matter density parameter.\footnote{This may introduce a mild internal inconsistency if the void model, which must be sourced by a cosmology with faster structure growth than $\Lambda$CDM, also modifies the background expansion.} The peculiar velocity $\vpec$ at the source position $\bm{r}$ is given by
\begin{equation}
    \vpec = \beta V_{\rm void}(\bm{r}) + \bm{V}_{\rm ext} \cdot \hat{\bm{r}},
    \label{eq:predicted_peculiar_velocity}
\end{equation}
where $V_{\rm void}(\bm{r})$ denotes the radial velocity predicted by the void model at position $\bm{r}$ in the reference frame comoving with the void. In~\cref{eq:predicted_peculiar_velocity}, we introduce $\beta$, a velocity scaling parameter that can either be treated as a free parameter or fixed to 1. By default, we treat it as a free parameter to mimic the effect of varying the void depth; it may also be interpreted simply as the average factor by which the model under- or over-predicts outflow velocities.

The \ac{TFR} relates the linewidth to the absolute magnitude $M$. Together with the apparent magnitude $m$, this determines the distance modulus $\mu$ as
\begin{equation}
    \label{eq:distmod_definition}
    \mu = m - M(\eta).
\end{equation}
The distance modulus is expressed as a function of the luminosity distance $d_{\rm L}$:
\begin{equation}
    \mu = 5 \log \frac{d_{\rm L}}{\mathrm{Mpc}} + 25,
\end{equation}
where the luminosity distance is related to the comoving distance $r$ by $d_{\rm L} = (1 + \zcosmo) r$.

We jointly infer the aforementioned void parameters and the distance indicator calibration, both collectively denoted as $\bm{\theta}$, along with the Bayesian evidence for each model. To achieve this, we use the \ac{TFR} flow model of~\citealt{VF_olympics} (hereafter~\citetalias{VF_olympics}), where we derive the likelihood of observing a galaxy with redshift $\zcmb$ (with uncertainty $\sigma_{cz}$/$c$) in the \ac{CMB} frame, apparent magnitude $\mobs$ (with uncertainty $\sigma_m$), and linewidth $\etaobs$ (with uncertainty $\sigma_\eta$), given the \ac{TFR} calibration and an underlying density and velocity field. In our case, these are the fields predicted by the void model.

For each galaxy, we introduce three latent parameters: its ``true'' comoving distance $r$, apparent magnitude $\mtrue$ at this distance, and linewidth $\etatrue$. All three are unknown and must be marginalised over. However, $r$ can be numerically marginalised during each likelihood evaluation, rather than directly sampled. The parameters $\mtrue$ and $\etatrue$ are related to the observed values through selection functions $S(\mobs)$ for apparent magnitude and $S(\etaobs)$ for linewidth, respectively, and the measurement uncertainties.

Together with the \ac{TFR} calibration, $\mtrue$ and $\etatrue$ provide an estimate of the galaxy distance modulus:
\begin{equation}
    \muTFR =
    \mtrue -
    \left\{
    \begin{array}{@{}l@{\;}l}
        \aTFR + \bTFR \etatrue + \cTFR \etatrue^2 & \text{if }\etatrue > 0, \\
        \aTFR + \bTFR \etatrue                    & \text{otherwise}.
    \end{array}
    \right.
    \label{eq:mu_TFR}
\end{equation}
The TFR-based $\muTFR$ is related to the true distance modulus $\mutrue$ through the intrinsic scatter of the \ac{TFR} and the homogeneous and inhomogeneous Malmquist biases. The homogeneous Malmquist bias is a simple $r^2$ volume factor in the prior on $r$, while the inhomogeneous Malmquist bias depends on the source number density $n(r)$. We assume that $n(r)$ is proportional to the matter density at the source position as predicted by the void model, such that $n(r) \propto \rho(r)$. In~\citetalias{VF_olympics}, we adopted a power-law bias of the form $\rho^\alpha$, treating $\alpha$ as a free parameter. In the main analysis here, we instead fix $\alpha = 1$ for simplicity. Nevertheless, we verify that allowing $\alpha$ to vary, or adopting a linear bias model, does not significantly affect our results. In particular, the void size constraints, presented later, change by at most 1 per cent under these alternative bias assumptions.
We use $r$ to calculate the cosmological redshift (Eq.~\ref{eq:redshift_to_distance}), which we combine with an estimate of the peculiar velocity at $r$ from the void model, thereby obtaining the predicted redshift $\zpred$ in the CMB frame (Eq.~\ref{eq:redshift_addition}). $\sigma_v$ is a Gaussian uncertainty between $c\zpred$ and $c\zcmb$, assumed to be position-independent. $\sigma_v$ captures the effect of small-scale motions that are not accounted for by the void model.

Thus, altogether the likelihood is as follows:
\begin{equation}
\label{eq:full_likelihood}
\begin{split}
    \mathcal{L}(\zcmb,\,\mobs,&\,\etaobs | \bm{\theta},\,\mtrue,\,\etatrue)
    =\\
    &\int \dd r \: \mathcal{N}\left(c\zcmb; c\zpred, \sqrt{\sigma_v^2 + \sigma_{c\zcmb}^2}\right)\\
    &\times\frac{r^2 n(r)\mathcal{N}\left(\mutrue; \muTFR, \sigmaTFR\right)}{\int {r^\prime}^2 n(r^\prime) \mathcal{N}\left(\mutrue^\prime; \muTFR, \sigmaTFR\right)\dd r^\prime} \\
    &\times \frac{S(\mobs) \mathcal{N}(\mobs; \mtrue, \sigma_m)}{\int S(\mobs^\prime) \mathcal{N}(\mobs^\prime; \mtrue, \sigma_m) \dd \mobs^\prime}\\
    &\times \frac{S(\etaobs) \mathcal{N}(\etaobs; \etatrue, \sigma_\eta)}{\int S(\etaobs^\prime) \mathcal{N}(\etaobs^\prime; \etatrue, \sigma_\eta) \dd \etatrue^\prime}.
\end{split}
\end{equation}
Further details are provided in~\citetalias{VF_olympics}. We adopt the same apparent magnitude selection as~\citetalias{VF_olympics}, which is modelled following~\citet{Boubel_2024}. Additionally, we assume that the selection in linewidth is a simple truncation, such that only values with $\etaobs > -0.3$ are observed. Assuming the galaxies are measured independently, the combined likelihood is given by the product of the likelihoods for individual galaxies.

Without additional information, the distribution of apparent magnitudes of sources at fixed absolute magnitude (or analogously $\etatrue$) is
\begin{equation}
    p(\mtrue | \etatrue,\,\bm{\theta}) \propto 10^{0.6 \, \mtrue}.
    \label{eq:prob_mag_true}
\end{equation}
This can be derived by assuming a uniform distribution of equal-magnitude objects with a radial distribution $\propto r^2$ in Euclidean space~\citep{Hubble_1926}. We sample $\etatrue$ from a Gaussian hyperprior as follows:
\begin{equation}
   p(\etatrue | \bm{\theta}) = \mathcal{N}(\etatrue | \hat{\eta},\,w_\eta),
\end{equation}
where $\hat{\eta}$ and $w_\eta$ denote the mean and standard deviation of the hyperprior, respectively. This is found to be a reliable model for the prior distribution of the latent parameters even when the true values are not Gaussian-distributed~\citep{MNR}.

We assign Jeffreys priors to $\sigma_v$ and $\sigmaTFR$, such that $\pi(\sigma_v) \propto 1 / \sigma_v$ and $\pi(\sigmaTFR) \propto 1 / \sigmaTFR$. For $\bm{V}_{\rm ext}$, we use a uniform prior in both magnitude and sky direction. We infer $\hat{\eta}$ with a uniform prior and $w_\eta$ with a Jeffreys prior. Uniform priors are also set for all remaining parameters. We summarize the model parameters in~\cref{tab:priors}. We sample the posterior using the \ac{HMC}-based No U-Turns Sampler~\citep[NUTS;][]{Hoffman_2011} as implemented in the \texttt{NumPyro} package~\citep{Phan_2019}.\footnote{\url{https://num.pyro.ai/en/latest/}} Burn-in samples are removed, and we ensure a sufficient number of steps such that the Gelman–Rubin statistic differs from unity by $<10^{-3}$~\citep{Gelman_1992}.

\begin{table*}
    \centering
    \begin{tabularx}{\textwidth}{>{\hsize=0.9\hsize}X >{\hsize=1.6\hsize}X >{\hsize=0.5\hsize}X}
        \toprule
        \textbf{Parameter} & \textbf{Description} & \textbf{Prior} \\[1mm]
        \multicolumn{3}{l}{\textit{Void parameters}} \\ \hline \hline
        $\RLG$              & Local Group offset from the centre of the void & Uniform \\
        $\Rvoid$            & Void size relative to the estimate of~\citetalias{Haslbauer_2020} & Uniform \\
        \multicolumn{3}{l}{\textit{Velocity field parameters}} \\ \hline \hline
        $\bm{V}_{\rm ext}$  & External velocity vector & Uniform in both magnitude and direction \\
        $\sigma_v$          & Scatter between the observed and predicted redshift & $\pi(\sigma_v) \propto 1 / \sigma_v$ \\
        $\beta$             & Velocity field scaling parameter & Uniform \\[3mm]
        \multicolumn{3}{l}{\textit{TFR distance calibration parameters}} \\ \hline \hline
        $\aTFR,\,\bTFR,\,\cTFR$ & TFR coefficients (zero-point, slope, and curvature) & Uniform \\
        $\sigmaTFR$ & TFR magnitude intrinsic scatter & $\pi(\sigmaTFR) \propto 1 / \sigmaTFR$ \\
        \multicolumn{3}{l}{\textit{Galaxy latent parameters}} \\ \hline \hline
        $\mtrue$ & Galaxy apparent magnitude & $p(\mtrue | \etatrue,\,\bm{\theta}) \propto 10^{0.6 \, \mtrue}$ \\
        $\etatrue$ & True galaxy linewidth & $\mathcal{N}(\etatrue | \hat{\eta}, w_{\eta})$ \\
        $\hat{\eta}$ & Mean of the linewidth Gaussian hyperprior & Uniform \\
        $w_{\eta}$ & Standard deviation of the linewidth Gaussian hyperprior & $\pi(w_{\eta}) \propto 1/w_{\eta}$\\
        \bottomrule
    \end{tabularx}
    \caption{Summary of the free parameters of our galaxy-by-galaxy flow model and their priors.}
    \label{tab:priors}
\end{table*}

We compute the model evidence $\mathcal{Z}$, which is defined as the integral of the product of the likelihood and prior over the parameter space.
\begin{equation}
    \mathcal{Z} \equiv \int \mathcal{L}(D | \bm{\bm{\theta}}) \pi(\bm{\theta}) \dd \bm{\bm{\theta}},
    \label{eq:evidence_definition}
\end{equation}
where $D$ is some data and $\bm{\theta}$ are the model parameters. The ratio of the evidences for two models, known as the Bayes factor $\mathcal{B}$, quantifies the relative support of the data for one model over another, assuming the two models are equally likely \textit{a priori}.

Since the flow model outlined above includes two latent variables per galaxy (the true magnitude and linewidth), the posterior of the full inference becomes $\approx14,000$-dimensional, rendering the evidence integral computationally infeasible. Consequently, whenever we calculate the evidence, we adopt the approximate flow model described in~\citetalias{VF_olympics}. In this approach, the true magnitude and linewidth are fixed to their observed values---effectively assigning them a Dirac-delta prior---while their uncertainties are linearly propagated into the distance modulus error. For a detailed discussion of this approximation, we refer the reader to Appendix~B of~\citetalias{VF_olympics}.

We calculate $\mathcal{Z}$ for the lower-dimensional posterior using \texttt{harmonic}, a tool that leverages normalising flows and the harmonic estimator to compute the evidence directly from posterior samples~\citep{McEwan_2021,Polanska_2024}.\footnote{\url{https://github.com/astro-informatics/harmonic}} To ensure that prior-range effects do not affect the evidence comparison, we adopt identical uniform parameter ranges when comparing different void profiles. When comparing the evidence of a void model to that of a model which assumes the velocity field is described solely by a constant dipole, the void model introduces two additional parameters with uniform priors relative to the dipole-only model: $\RLG$, with a uniform prior ranging from $-50$ to 50~Mpc; and $\tilde{r}_{\rm void}$, with a uniform prior ranging between 0.01 and 3. These two parameters reduce the void model evidence by approximately 2.48 in $\log_{10}$ compared to the fiducial void model, in which they are fixed.


\section{Results}
\label{sec:results}

In this section, we present the results of calibrating the void model parameters using both the bulk flow curve from~\citetalias{Watkins_2023} and the galaxy-by-galaxy flow model applied to the \ac{CF4} \ac{TFR} galaxy sample with the SH0ES Cepheid absolute calibration. For the latter, we compute the Bayesian evidence to compare different void profiles and identify which is most favoured by the data (\cref{sec:results_model_selection}). Additionally, in~\cref{sec:results_fiducial}, we examine the special case where the void size is fixed to its fiducial value, presenting corrected results for the~\citetalias{Mazurenko_2024} analysis.

\subsection{Galaxy-by-galaxy constraint}\label{sec:results_varying}

\subsubsection{Inferred void parameters}

\begin{figure*}
    \centering
    \includegraphics[width=\textwidth]{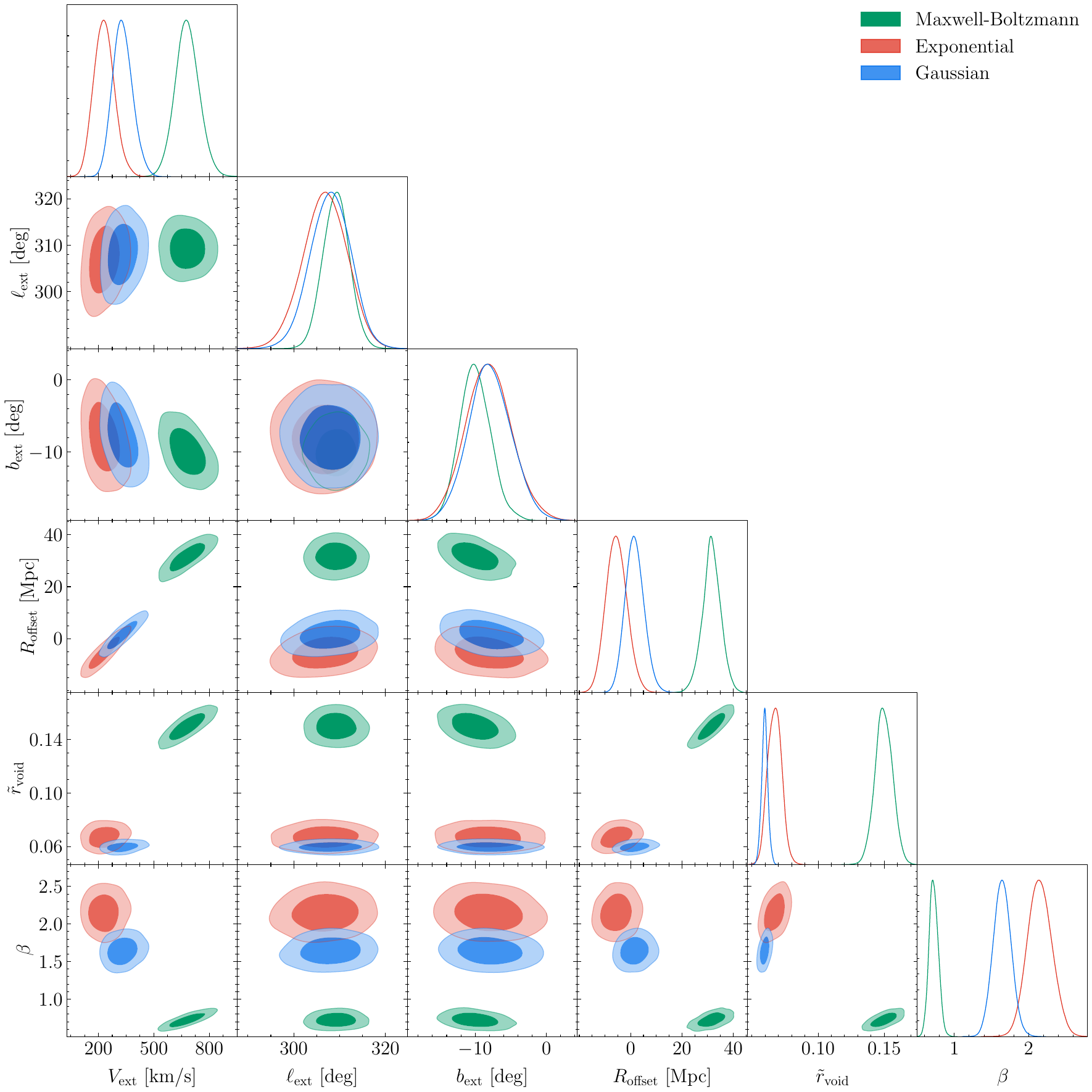}
    \caption{Posterior distributions of the inferred void model parameters obtained by fitting the flow model described in~\cref{sec:flow_model} to the \ac{CF4} \ac{TFR} data. Results are shown for the exponential (red), Gaussian (blue), and Maxwell-Boltzmann (green) void profiles. The sampled parameters include the magnitude and direction of the external velocity $\bm{V}_{\rm ext}$ in Galactic coordinates, the observer's offset $\RLG$ from the void centre along the symmetry axis, the relative void size $\tilde{r}_{\rm void}$, and the velocity scaling parameter $\beta$. For all three profiles, we find a preference for significantly smaller void sizes compared to the fiducial sizes in~\citetalias{Haslbauer_2020}.}
    \label{fig:extended_void}
\end{figure*}

In~\cref{fig:extended_void}, we present the results of jointly inferring all model parameters (summarized in \cref{tab:priors}) for the three void profiles, following the methodology described in~\cref{sec:flow_model}. For all profiles, we find that $\bm{V}_{\rm ext}$ points towards a consistent direction of $(\ell_{\rm ext}, \, b_{\rm ext}) \approx (308^\circ,\, -8^\circ)$ in Galactic coordinates. However, its magnitude varies significantly across the profiles, with best-fitting values of 230, 331, and 681~km/s for the exponential, Gaussian, and Maxwell-Boltzmann cases, respectively, each with a typical uncertainty of $\approx 50$~km/s.

As expected, the magnitude of $\bm{V}_{\rm ext}$ is degenerate with the observer offset $\RLG$, for which we infer values of $-5.5 \pm 4$, $1.5 \pm 3.6$, and $31.5 \pm 3.6$~Mpc for the exponential, Gaussian, and Maxwell-Boltzmann profile, respectively. While the exponential and Gaussian profiles favour a position near the void centre, the Maxwell-Boltzmann case requires a significantly offset observer. As we will show in~\cref{fig:extended_void_stats}, the large $\bm{V}_{\rm ext}$ effectively counteracts the larger void-induced outflow at $\RLG$, since the outflow vanishes at the void centre and rises roughly linearly at small radii.

The most surprising feature of~\cref{fig:extended_void} is the very low inferred fractional void size $\tilde{r}_{\rm void}$, with values of $0.0670 \pm 0.0052$, $0.0597 \pm 0.0024$, and $0.1500 \pm 0.0066$ for the exponential, Gaussian, and Maxwell–Boltzmann profile, respectively. These correspond to preferred initial void sizes of approximately $69$, $61$, and $34~\mathrm{cMpc}$, substantially smaller than the fiducial values used in~\citetalias{Haslbauer_2020}. Such small voids produce no significant imprint on the CMB quadrupole, making the CMB quadrupole constraint irrelevant.

Lastly, we find that the velocity scaling parameter $\beta$ is inferred to be $2.15 \pm 0.16$, $1.64 \pm 0.12$, and $0.72 \pm 0.06$ for the exponential, Gaussian, and Maxwell-Boltzmann profiles, respectively. None of the profiles are consistent with $\beta = 1$, indicating that the fiducial void-predicted velocities require rescaling. Specifically, the exponential and Gaussian profiles underestimate the outflow velocities (for the best-fit small $\tilde{r}_{\rm void}$), while the Maxwell-Boltzmann profile overestimates them. For the exponential and Gaussian profiles, these results suggest a smaller but deeper void than reported in~\citetalias{Haslbauer_2020}, an issue we investigate next.

\subsubsection{Inferred void density and velocity profiles}

\begin{figure*}
    \centering
    \includegraphics[width=\textwidth]{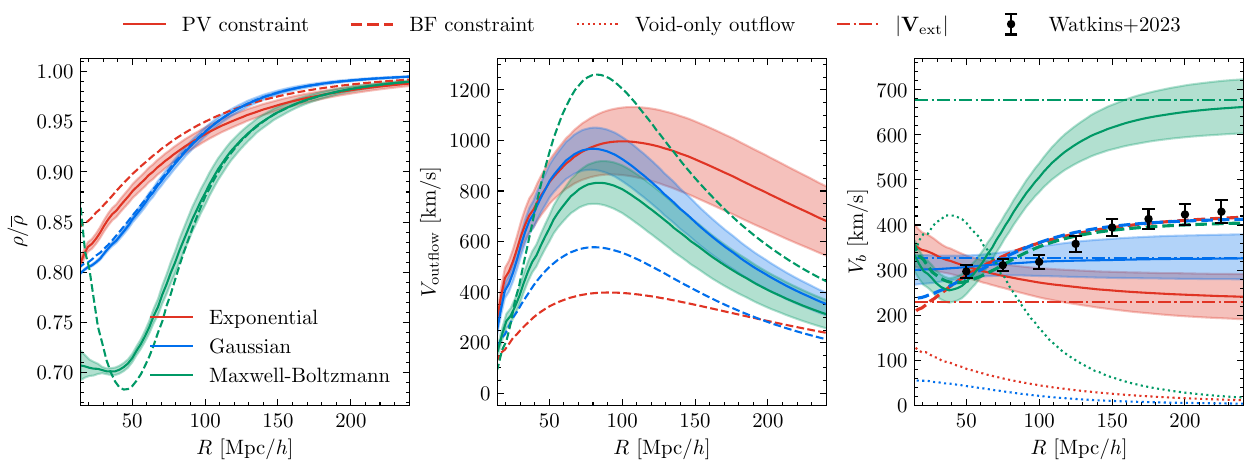}
    \caption{The inferred void density field (left panel), the outflow velocity curve (i.e., the monopole of the velocity field; middle panel), and the bulk flow curve (an integral of the dipole of the velocity field; right panel) as a function of distance from the observer, using the void parameters and their uncertainties as inferred by our field-level analysis. We show results for the exponential (Gaussian; Maxwell-Boltzmann) profile using red (blue; green) shaded bands, which indicate the $1\sigma$ uncertainty. In the right panel, the dot-dashed lines show $V_{\rm ext}$ and the dotted lines show the bulk flow without including $\bm{V}_{\rm ext}$, i.e., that produced intrinsically by the void. The dashed lines show the bulk flow curves in the models that best fit the results of~\protect\citetalias{Watkins_2023}, where the colour again indicates the density profile. The~\protect\citetalias{Watkins_2023} results themselves are shown as black points with uncertainties.}
    \label{fig:extended_void_stats}
\end{figure*}

A much smaller local void implies a more rapid rise in density towards the cosmic mean, as shown in the left panel of~\cref{fig:extended_void_stats}. This density profile is calculated from the final positions of the particle trajectories using Equations~54 and 55 of~\citetalias{Haslbauer_2020} for the Maxwell-Boltzmann profile, which we generalize appropriately for the other void profiles. The outflow velocity curve is computed as the mean enclosed radial velocity, whereas the bulk flow curve is computed as the mean enclosed 3D velocity. The middle panel of~\cref{fig:extended_void_stats} illustrates that while the void-induced outflow velocity $V_{\rm outflow}$ increases at a rate comparable to the Hubble tension ($cz' - \dot{a}$), it reaches its peak at significantly smaller radii than those used to measure $cz'$ due to the more compact void profiles.

The right panel of~\cref{fig:extended_void_stats} shows the reconstructed bulk flow curves. None of the preferred models match the results of~\citetalias{Watkins_2023}. The exponential and Gaussian profiles yield nearly flat bulk flow curves dominated by $V_{\rm ext}$, as expected given that the inferred observer position is very close to the void centre. In contrast, the Maxwell-Boltzmann profile produces a steeply rising bulk flow curve, driven by the relatively large $V_{\rm ext}$, which sets the asymptotic limit. However, the rapid rise with radius overshoots even the steep bulk flow trend reported by~\citetalias{Watkins_2023}. We discuss this further in~\cref{sec:ramification}. It is important to bear in mind that these posterior predictive bulk flows are derived through the lens of the void model and do not represent a general Bayesian inference of the bulk flow from the peculiar velocity data. They do however suggest that such an inference would not agree with the~\citetalias{Watkins_2023} result. The void models predict a bulk flow of $\approx 300~\mathrm{km}/\mathrm{s}$ at the origin, where we also have the constraint that the Local Group motion with respect to the CMB is $620\pm15~\mathrm{km}/\mathrm{s}$~\citep{Planck_2018}. These are consistent given the velocity dispersion parameter $\sigma_v \approx 350~\mathrm{km}/\mathrm{s}$, which describes the scatter between the motion of individual tracers (such as a halo hosting the \ac{LG}) and the smooth large-scale velocity field captured by the void models.

\subsubsection{Predicted local $H_0$}

Although the velocity data does appear to favour the existence of a void, the fact that it is much smaller than the~\citetalias{Haslbauer_2020} model means that the outflow velocity is too low and peaks too quickly to solve the Hubble tension. Local measurements of $cz'$ typically correspond to distances of $100-600$~Mpc~\citep[$z = 0.023-0.15$;][]{Camarena_2020a}, well beyond the outflow peak. We quantify this by computing the Hubble constant $H_0^{\rm local} \equiv cz'$ that would be inferred locally in these models, following the method described in Section~3.3.4 of~\citetalias{Haslbauer_2020}. We note that the actual value of $H_0$, or the present $\dot{a}$, is 67.4~km/s/Mpc in all models.

\begin{figure}
    \centering
    \includegraphics[width=\columnwidth]{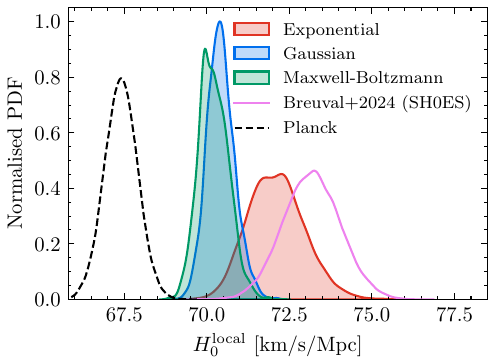}
    \caption{Posterior predictive distributions of $H_0^{\rm local}$ when jointly varying the void size and velocity scaling parameter $\beta$ for the exponential, Gaussian, and Maxwell-Boltzmann void profiles (shown in red, blue, and green respectively). We find $H_0^{\rm local} = 72.08^{+0.86}_{-0.80}$, $70.40^{+0.43}_{-0.38}$, and $70.18^{+0.48}_{-0.40}$~km/s/Mpc for the exponential, Gaussian, and Maxwell-Boltzmann profiles, respectively. For comparison, we show the \emph{Planck} and 4-anchor SH0ES $H_0$~\protect\citep{Planck_2020, Breuval_2024} as unfilled dashed black and solid magenta distributions, respectively. The \emph{Planck} $H_0$ is the background value assumed by the void models.}
    \label{fig:extended_void_H0}
\end{figure}

In~\cref{fig:extended_void_H0}, we show that the very small inferred void sizes limit the ability of the Gaussian and Maxwell-Boltzmann profiles to fully resolve the Hubble tension. Specifically, they yield $H_0^{\rm local} = 70.40^{+0.43}_{-0.38}$ and $70.18^{+0.48}_{-0.40}$~km/s/Mpc, respectively -- about $3\sigma$ below the 4-anchor SH0ES measurement of $73.17 \pm 0.86$~km/s/Mpc~\citep{Breuval_2024}. In contrast, the exponential void profile remains a plausible solution, producing $H_0^{\rm local} = 72.08^{+0.86}_{-0.80}$~km/s/Mpc, which is consistent with local determinations~\citep{Uddin_2024}. In fact, this $H_0^{\rm local}$ is only $1\sigma$ below the SH0ES measurement and only $1-2\sigma$ below the slightly higher more recent measurements with little or no reliance on SNe~Ia~\citep{Boubel_2024, Jensen_2025, Scolnic_2025, Vogl_2025}. We note that in these studies, the top rung of the distance ladder uses techniques that do not reach out as far as SNe~Ia, so $H_0^{\rm local} \equiv cz'$ has to be estimated using a more limited redshift range~\citep[e.g., $z = 0.023 - 0.1$ for][]{Scolnic_2025}. Given the small void sizes inferred in our analysis, working out the local $cz'$ over a narrower redshift range would most probably slightly increase the model value, further reducing the tension.

\subsubsection{Model selection}\label{sec:results_model_selection}

We now compare the performance of the void models using a Bayesian evidence analysis~\citep{Jeffreys_1939}. The resulting $\log \mathcal{B}$ is 10.7 (9.4) for the Gaussian (Maxwell-Boltzmann) profile relative to the result for the exponential profile. The positive values in both cases show that the exponential profile is strongly disfavoured compared to the other profiles. However, the difference between the Gaussian and Maxwell-Boltzmann profiles is modest, indicating only weak preference for the Gaussian model despite the markedly different preferred value of $\bm{V}_{\rm ext}$. We note that these Bayesian evidences do not take into account measurements of the local $cz'$, which we saw in the previous section clearly prefers the Exponential model~(\cref{fig:extended_void_H0}).

To better understand the impact of introducing the velocity scaling parameter $\beta$, we also consider models with fixed $\beta = 1$. In this case, we find $\log \mathcal{B} = 17.3$ (22.7) for the Gaussian (Maxwell-Boltzmann) profile relative to the exponential profile result. This implies a strong preference for the Maxwell-Boltzmann model if $\beta = 1$. Notably, however, the evidence for the Maxwell-Boltzmann profile remains nearly unchanged with or without $\beta$, while the exponential and Gaussian models gain significantly from allowing $\beta$ to vary, improving by approximately 13 and 6 in $\log \mathcal{B}$, respectively. This suggests that jointly varying both the relative void size and the velocity scaling parameter (effectively analogous to the initial void underdensity) is essential for a good fit to the distance and redshift data, except in the case of the Maxwell-Boltzmann profile.

\begin{figure}
    \centering
    \includegraphics[width=\columnwidth]{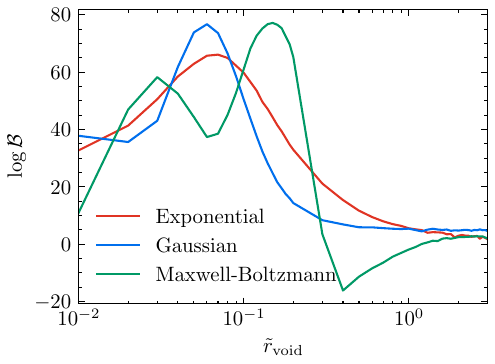}
    \caption{The logarithmic Bayes factor for a grid of void sizes, which we hold fixed during each inference. The Bayes factors are relative to a model where the velocity field is modelled only by a constant dipole $\bm{V}_{\rm ext}$, with positive values indicating a preference over this simple void-free model. While most of the probed range of void sizes is preferred over the simpler pure dipole model with no net outflow, the preference is much stronger for voids significantly smaller than the fiducial size found by~\citetalias{Haslbauer_2020}. The two preferred models are the Gaussian and Maxwell-Boltzmann profiles with $\tilde{r}_{\rm void} \approx 0.06$ and $0.15$, respectively.}
    \label{fig:void_size_evidence}
\end{figure}

To investigate this further, we compute $\mathcal{B}$ for a grid of void sizes, holding $\tilde{r}_{\rm void}$ fixed each time (the ``profile Bayes factor''). The scaled void radius is increased in steps of 0.01 up to $\tilde{r}_{\rm void} = 0.2$, and in steps of 0.1 beyond that. Our results are presented in~\cref{fig:void_size_evidence}, where we show the Bayes factors relative to a model in which the velocity field is modelled solely by $\bm{V}_{\rm ext}$. Despite the higher complexity of the void models, they are strongly preferred by the data over a model with fixed $\bm{V}_{\rm pec} = \bm{V}_{\rm ext}$ throughout the considered volume for most considered values of $\tilde{r}_{\mathrm{void}}$. For the exponential and Gaussian profiles, this is true even for the fiducial void size found by~\citetalias{Haslbauer_2020}. As expected, the exponential, Gaussian, and Maxwell-Boltzmann profiles exhibit maxima near $\tilde{r}_{\rm void} \approx 0.07$, $0.06$, and $0.15$, respectively. For smaller void sizes, the underdense region becomes negligible, leading to virtually no outflow. \cref{fig:void_size_evidence} also shows that larger void sizes, particularly the fiducial size proposed by~\citetalias{Haslbauer_2020}, are disfavoured by the data compared to smaller voids, which provide a better fit. It will be interesting to explore the extent to which such a small void would be consistent with other datasets, but this goes beyond the scope of the present work.

\subsubsection{Distance to the Coma Cluster}
\label{sec:distance_to_coma}

As a posterior predictive check of the void models, we examine their predictions for the redshift of the Coma Cluster, one of the most massive and extensively studied nearby galaxy clusters~\citep[e.g.][]{Peebles_1970, Colless_1996, Pimbblet_2014}. Recently,~\citet{Scolnic_2025} used a calibration of the SNe~Ia absolute magnitude based on the \textit{HST} distance ladder to estimate a distance to Coma of $98.5 \pm 2.2$~Mpc, along with a CMB-frame observed redshift of $0.02445 \pm 0.00024$. This is consistent with other recent estimates such as~\citet{Benisty_2025}, who report a heliocentric redshift of $0.02333 \pm 0.00013$, corresponding to $0.02424 \pm 0.00013$ in the CMB frame.~\citet{Scolnic_2025} emphasized that this distance is substantially lower than the value inferred from the \emph{Planck} cosmology assuming negligible peculiar velocity. Reconciling the observed redshift with the calibrated distance would require a mean peculiar velocity exceeding $500~\mathrm{km}/\mathrm{s}$, despite no velocity reconstructions predicting an abnormally large peculiar velocity for Coma~\citep[e.g.][]{Carrick_2015, Jasche_2019, Lilow_2021}.

\begin{figure*}
    \centering
    \includegraphics[width=\textwidth]{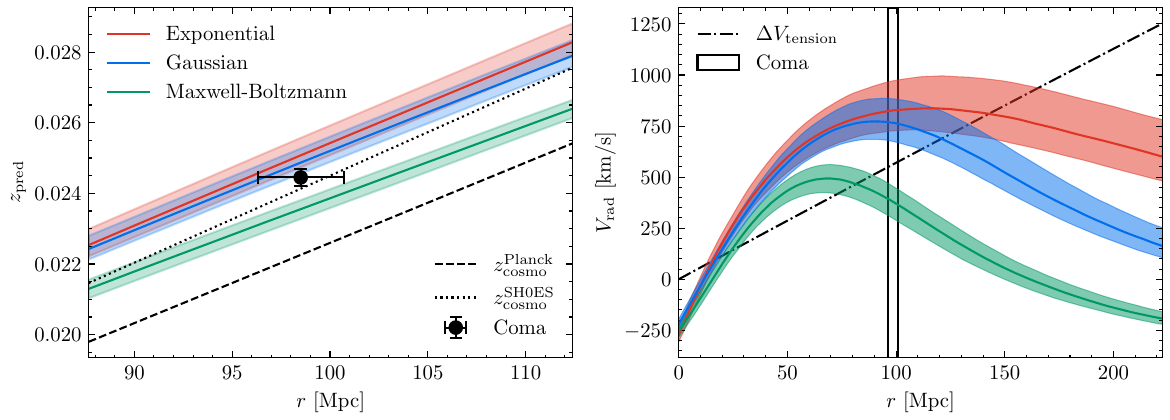}
    \caption{Predicted redshift of the Coma cluster as a function of its comoving distance $r$ (left panel) and the corresponding radial velocity profile (right panel), shown for the three void profiles. Shaded bands indicate the $1\sigma$ uncertainty regions. For the Coma cluster, we assume $r = 98.5 \pm 2.2$~Mpc and $\zcmb = 0.02445 \pm 0.00024$~\citep{Scolnic_2025}, both shown with $1\sigma$ error bars. The dashed and dotted black lines in the left panel show the redshift-distance relation for the \emph{Planck} and SH0ES values of $H_0$, respectively, assuming no peculiar velocity in either case. In the right panel, the vertical band indicates the $1\sigma$ distance to Coma from the \textit{HST} distance ladder calibration, while the dash-dotted line shows the peculiar velocity required in the \emph{Planck} cosmology to reproduce the same total redshift at a given physical distance as in the SH0ES cosmology with no peculiar velocity ($\Delta V_{\rm tension}$). The observed redshift and distance of Coma imply a slightly higher local $cz'$ than in the SH0ES cosmology~\citep{Scolnic_2025}. Given the observed redshift of Coma, both the exponential and Gaussian profiles yield predictions consistent with its observed distance, which the Maxwell-Boltzmann profile mildly overestimates at the $2\sigma$ level.}
    \label{fig:coma_distance}
\end{figure*}

The void models we consider naturally produce such outflows while assuming the \emph{Planck} $H_0$ at the background level. The left panel of~\cref{fig:coma_distance} shows the predicted redshift $\zpred$ as a function of comoving distance $r$ to Coma, while the right panel shows the corresponding radial velocity profile. Among the models, the Gaussian profile shows the best agreement with the calibrated Coma distance, though the exponential profile yields a prediction that is only marginally different. In contrast, the Maxwell-Boltzmann profile, due to its weaker outflow, over-predicts the distance to Coma at the $2\sigma$ level.

\subsection{Bulk flow constraints}\label{sec:res_bulk_flow_size}

We now perform a $\chi^2$ analysis by fitting the void model to the bulk flow curve of~\citetalias{Watkins_2023} (\cref{sec:bf_model}). This will allow us to correct the factor-of-three error described in Sec.~\ref{sec:bf_model}, while updating the constraints according to our field-level findings. For computational efficiency, we fix the relative void sizes to the best-fitting values from the previous section: 7, 6, and 15 per cent for the exponential, Gaussian, and Maxwell-Boltzmann profile, respectively. We sample the observer offset $\RLG$ and external velocity $\bm{V}_{\rm ext}$, assuming its direction is aligned with the previously inferred best-fit direction from the galaxy-by-galaxy analysis. The goal is to assess whether these more compact void profiles still allow for a satisfactory fit to the observed bulk flow.

\begin{figure}
    \centering
    \includegraphics[width=\columnwidth]{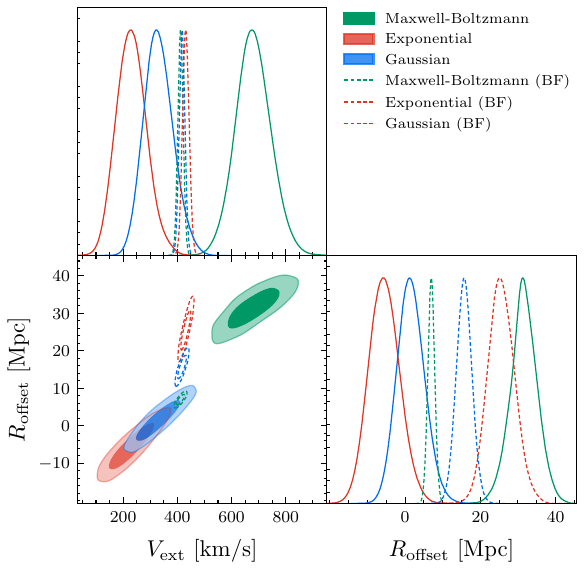}
    \caption{Comparison of the void systemic velocity magnitude $V_{\rm ext}$ and the observer offset $\RLG$ between the field-level analysis (based on the~\ac{CF4} catalogue) and the bulk flow constraints (``BF'', measured by~\protect\citetalias{Watkins_2023} but also based on~\ac{CF4}). The discrepancy casts doubt on their bulk flow inference as the underlying data is the same in both cases.}
    \label{fig:void_size_comparison}
\end{figure}

In~\cref{fig:void_size_comparison}, we compare the results of this fit to those from the galaxy-by-galaxy flow model. We find that even such relatively small voids can reproduce the~\citetalias{Watkins_2023} bulk flow curve well. For the exponential profile, the best fit is obtained at $\RLG = 25$~Mpc and $V_{\rm ext} = 430$~km/s, with a minimum $\chi^2 = 9.14$. The Gaussian profile yields a best fit at $\RLG = 16~\mathrm{Mpc}$ and $V_{\rm ext} = 419$~km/s, with $\chi^2 = 9.40$. Lastly, the Maxwell-Boltzmann profile yields a best fit at $\RLG = 7$~Mpc and $V_{\rm ext} = 412$~km/s, with $\chi^2 = 11.01$. All these $\chi^2$ values are very reasonable for 9 data points and 2 free parameters.

However, these best-fit parameters are in significant tension with those inferred from the flow model analysis, which favours substantially lower values of both $V_{\rm ext}$ and $\RLG$. This discrepancy is a manifestation of the fact that the flow model prefers nearly flat or even declining bulk flow curves for the exponential and Gaussian profiles (Fig.~\ref{fig:extended_void_stats}), while fitting directly to the bulk flow curve forces the model to produce an increasing trend with radius, pushing the fit toward higher $V_{\rm ext}$ and larger offsets. The other unexpected feature is that the uncertainties appear smaller in the bulk flow fit. This may suggest that the error bars reported by~\citetalias{Watkins_2023} are underestimated~\citep[as noted by][]{Whitford_2023}, or it could reflect the simplifying assumption in these constraints that the bulk flow measurements are statistically independent.

Lastly, to connect with the analysis of~\citetalias{Mazurenko_2024}, we present updated results using the corrected definition while keeping the void size fixed to its fiducial value (\cref{sec:results_fiducial_bf}). We find that the fiducial size void profiles remain capable of reproducing the bulk flow curve measured by~\citetalias{Watkins_2023}, although they now favour different values of $\bm{V}_{\rm ext}$ and $\RLG$ compared to the original analysis. This suggests that the~\citetalias{Watkins_2023} bulk flow curve can be reproduced with little difficulty by the considered void models, which is not the case in $\Lambda$CDM (see their Figure~8).

\section{Discussion}\label{sec:discussion}

\subsection{Implications and significance of the results}\label{sec:ramification}

For the first time, we test the local void scenario for the Hubble tension~\citepalias{Haslbauer_2020} and constrain its degrees of freedom by comparing with direct distances to individual galaxies, which combined with their redshifts tells us their peculiar velocity. In particular, we evaluate the void models directly against the relevant galaxy observables (sky position, redshift, linewidth, and apparent magnitude) using the field-level flow model described in~\cref{sec:flow_model}, while jointly inferring the parameters of the \ac{TFR} used to estimate redshift-independent distances from the linewidth (Eq.~\ref{eq:eta_definition}). Our models assume a background \emph{Planck} cosmology that fits the CMB anisotropies~\citep{Planck_2020, Tristram_2024, Calabrese_2025, Camphuis_2025}.

For the fiducial void size, our field-level inference of $\bm{V}_{\rm ext}$ and $\RLG$ yields results consistent at $<2\sigma$ with those obtained from fitting to the~\citetalias{Watkins_2023} bulk flow curve under the assumption of independent data points (\cref{fig:fiducial_void}). Moreover, we find excellent agreement between the bulk flow direction inferred from the \ac{CF4} dataset and that reported by~\citetalias{Watkins_2023}. However, we find significantly different results when we relax the assumption of a fixed void size and allow for an effect similar to variation in the initial void depth, captured through a velocity scaling parameter $\beta$. In these extended models, our field-level analysis implies bulk flow curves dominated by the external velocity $\bm{V}_{\rm ext}$. With the exception of the poorly-fitting Maxwell-Boltzmann void profile, the posterior predicted bulk flow curves do not rise, in disagreement with the result of~\citetalias{Watkins_2023}. This is potentially related to the findings of~\citet{Whitford_2023} that the uncertainties in the bulk flow estimator are likely larger than those quoted by~\citetalias{Watkins_2023}, particularly the latter's claim of achieving 37~km/s precision out to $200~\mathrm{Mpc} / h$.

The void scenario was previously tested by~\citetalias{Mazurenko_2024} using the bulk flow curve from~\citetalias{Watkins_2023}. However, their analysis contained a significant flaw in the definition of the bulk flow (see Section~\ref{sec:bf_model}). Adopting the correct formulation and generalizing the~\citetalias{Haslbauer_2020} model to include an arbitrary systemic void velocity $\bm{V}_{\rm ext}$ in addition to the void outflow, we find that a good fit to the~\citetalias{Watkins_2023} bulk flow curve is possible, provided the observer lies within $\la 50$~Mpc of the void centre (as discussed in~\cref{sec:results_fiducial}). However, this agreement no longer constitutes an \emph{a priori} prediction of the void model. Rather, it highlights that the generalisation of the void model's systemic velocity allows for the reproduction of a broad range of bulk flow curves, including that of \citetalias{Watkins_2023}. This is because the void scenario generically assumes enhanced growth of structure compared to $\Lambda$CDM on $\ga 100$~Mpc scales, which aligns well with the implications of the~\citetalias{Watkins_2023} result.

Crucially, our main analysis shows that the \ac{CF4} peculiar velocity data prefers significantly smaller void sizes for all profiles compared to the fiducial void size found by~\citetalias{Haslbauer_2020}. While our approach tests the predicted velocity field on a galaxy-by-galaxy basis,~\citetalias{Haslbauer_2020} constrain the void using source number densities and its ability to resolve the Hubble tension. The two methods are complementary but distinct, and their discrepancy may point either to limitations of the KBC void model itself---such as some systematic bias---or to the need for a more complex void density profile. It is interesting to note that when calibrating the void model to match the bulk flow curve of~\citetalias{Watkins_2023}, the $\chi^2$ comparison favours the fiducial void size over the smaller sizes preferred by the field-level analysis (\cref{sec:results_fiducial_bf}).

While the bulk flow is a useful summary statistic, there are uncertainties when estimating it from the \ac{CF4} survey~\citep{Tully_2023}. \ac{CF4} combines several dedicated surveys with varying sky coverage, which implies that any magnitude mis-calibration could introduce spurious bulk flows that increase with distance. This is due to the gradient of redshift with respect to distance modulus, $\dd (cz_{\rm cosmo}) / \dd \mu$. Consequently, a zero-point calibration offset as small as $0.017~\mathrm{mag}$---corresponding to the statistical $1\sigma$ uncertainty in the calibration of the \ac{TFR} or fundamental plane zero-point relative to SNe~Ia in the \ac{CF4} catalogue~\citep[see Figure~16 of][]{Tully_2023}---would induce an artificial flow of 76 (152)~km/s at 100 (200)~Mpc$/h$. This concern is particularly relevant for \ac{CF4}, where the sources beyond $150~\mathrm{Mpc}/h$ come almost exclusively from the \ac{SDSS} fundamental plane dataset. Another potential concern is that the methodology of~\citetalias{Watkins_2023} underestimates uncertainties by a factor of $\approx 3.5$, with this discrepancy increasing at larger distances, precisely where~\citetalias{Watkins_2023} reports the strongest tension with $\Lambda$CDM~\citep{Whitford_2023}.

Using a Bayesian evidence comparison, as presented in Fig.~\ref{fig:void_size_evidence}, we find that the Gaussian profile is marginally preferred over the Maxwell-Boltzmann profile, while the exponential profile is decisively disfavoured. This outcome is somewhat surprising given that the Gaussian and Maxwell-Boltzmann profiles favour significantly different values of $\bm{V}_{\rm ext}$. In the Maxwell-Boltzmann case, the predicted bulk flow magnitude approaches $\approx 700$~km/s at $R > 200~\mathrm{Mpc}/h$ because $\bm{V}_b \to \bm{V}_{\rm ext}$ at large radii due to the decay of the void-induced outflow. However, taking this extrapolation at face value is problematic, as the majority of the data lies at much smaller radii: the \ac{CF4} galaxy distribution peaks near $60~\mathrm{Mpc}/h$ and drops off sharply beyond $120~\mathrm{Mpc}/h$ (\cref{fig:CF4_redshift_sky_dist}). At the peak, the large $\bm{V}_{\rm ext}$ acts to cancel the void-induced outflow due to the significantly offset observer position, which may dominate the likelihood. This comes at the cost of over-predicting the bulk flow for the relatively few galaxies located beyond $100~\mathrm{Mpc}/h$, since the assumption of a constant external flow is clearly too simplistic. Despite these difficulties, the exponential and Gaussian void models are preferred over a simpler $\bm{V}_{\rm ext}$-only model with no void outflow and thus effectively no void, for all void sizes (see~\cref{fig:void_size_evidence}).

Although a small void can reproduce the~\citetalias{Watkins_2023} bulk flow curve if calibrated to it (see the right panel of~\cref{fig:extended_void_stats}), the preferred parameters are discrepant with those based on the field-level analysis of either profile. Furthermore, out of the smaller void models preferred by our field-level analysis, only the exponential profile produces a sufficiently large outflow to match the observed ${H_0^{\rm local} \equiv cz'}$ at just over $1\sigma$. However, it is also the least favoured profile according to our field level analysis with free void size. This instead favours the Gaussian and Maxwell-Boltzmann profiles, which however still leave a residual $3\sigma$ Hubble tension as they are only able to account for about half of it. This limitation arises because the outflow velocity peaks at low radii, leading to only a modest increase in $cz'$ over the redshift range commonly used to infer it~\citep[${z = 0.023-0.15}$, e.g.][]{Riess_2022_comprehensive}. The Hubble tension can be fully resolved with the fiducial void sizes considered by \citetalias{Haslbauer_2020}, which however are larger than the preferred sizes inferred here. Nonetheless, these results demonstrate that (for the data we use) a local void remains a viable mechanism for alleviating the Hubble tension, even if not fully resolving it. Note that our field-level analysis could conceivably have inferred $\beta < 0$ and placed us in an overdensity, since only velocity data is used in the fit. Moreover, the Bayesian evidence could have favoured the simpler $\bm{V}_{\rm ext}$-only model.

It is interesting to consider why our field-level analysis consistently favours such a small void size when this parameter is allowed to vary, especially if one accepts the presence of a large-scale underdensity in the luminosity density extending out to 300~Mpc, as posited in the KBC void scenario upon which the fiducial void sizes were originally based~\citep{Keenan_2013}. One possibility is that this could hint at a generally more complex void profile, e.g. a `void-in-void' configuration, whereby the KBC void encompasses a smaller void that contains the \ac{LG}. A more complex structure, which necessitates the inclusion of an accurate description of smaller scales, implies a departure from just a spherically symmetric smooth large-scale void, as is described by the~\citetalias{Haslbauer_2020} models. The void-in-void scenario is disfavoured by our analysis given the absence of any secondary features in $\log \mathcal{B}$ at void sizes comparable to that inferred by ~\citetalias{Haslbauer_2020} (see Fig.~\ref{fig:void_size_evidence}), at least for the exponential and Gaussian profiles, which we consider more promising on other grounds (see Fig.~\ref{fig:extended_void_stats}). Importantly, given that the \ac{CF4} catalogue extends robustly only to $\approx 120$~Mpc (Fig.~\ref{fig:CF4_redshift_sky_dist}) and becomes increasingly sparse beyond that, the outer regions of any larger-scale underdensity are only weakly probed. Thus, the velocity imprint of a more extended, shallower component of the void could plausibly evade detection in the present analysis. A more sophisticated treatment would likely be required to capture the full dynamics of such a configuration.

Another possibility is simply that the Hubble tension should be solved with a smaller and most likely deeper void than inferred by~\citetalias{Haslbauer_2020}. We note that part of the reason~\citetalias{Haslbauer_2020} preferred such large void sizes is to fit the high $H_0$ values from H0LiCOW strong lensing time delays~\citep{Wong_2020}. Subsequent updates to this analysis indicate good agreement with $H_0^{\mathrm{Planck}}$~\citep{Birrer_2020, Li_2025}, as is also the case with strong lensing time delays of SN Refsdal~\citep{Kelly_2023, Grillo_2024} and CC results at similar redshifts~\citep{Cogato_2024, Guo_2025}. This calls for a more rapid decay to the effects induced by a local void. Since we have already seen that an exponential void with a much smaller size than inferred by~\citetalias{Haslbauer_2020} can adequately explain the locally measured $cz'$ (\cref{fig:extended_void_H0}) and the \ac{CF4} galaxy data at the field level, it is worthwhile to explore if it can fit other observations and yield redshifts at fixed distance that converge at the correct rate to those in the background \emph{Planck} cosmology~\citep{Mazurenko_2025}. We note that since the galaxy number counts of~\citet{Keenan_2013} are in redshift rather than real space and the effect of a void on redshifts would persist even after the density has returned to the cosmic mean, it is conceivable that their reported luminosity underdensity can be recovered by the smaller voids proposed here (especially if the void is deeper than considered by~\citetalias{Haslbauer_2020}).

As described in~\cref{sec:data}, we use an absolute calibration of the \ac{CF4} \ac{TFR} using 12 galaxies with Cepheid distances measured as part of the SH0ES project. This partially breaks the degeneracy between $H_0$ and the normalization of the \ac{TFR}, making our analysis somewhat sensitive to $H_0$ directly. We find that the absolute calibration has virtually no effect on the inferred void size, with the preferred value shifting by $\la 1$ per cent. Likewise, neither $V_{\rm ext}$ nor $\RLG$ change appreciably for the exponential or Gaussian profiles. Only in the case of the Maxwell–Boltzmann profile does $V_{\rm ext}$ increase, reaching $\approx 900$~km/s when no absolute calibration is applied and $\beta = 1$.

In contrast, the inclusion of the absolute calibration significantly impacts $\beta$: without it, $\beta$ remains unconstrained due to degeneracies with the void outflow and the \ac{TFR} normalization. These degeneracies are only broken when an absolute calibration is imposed. In our analysis, the local Hubble constant depends on the void size and the velocity scaling parameter $\beta$ as
\begin{equation}
    H_0^{\rm local}(\beta) = H_0^{\rm Planck} + \beta \left(\left. H_0^{\rm local} \right|_{\beta = 1} - H_0^{\rm Planck}\right),
\end{equation}
as $\beta$ scales the deviations from the background \emph{Planck} cosmology and the void size enters only through $\left.H_0^{\rm local}\right|_{\beta = 1}$, the predicted $H_0^{\rm local}$ when $\beta = 1$. Since the inferred void size is unaffected by the absolute calibration, in practice the calibration affects $H_0^{\rm local}$ only through $\beta$. If $\beta$ were fixed to unity and no absolute calibration applied, the resulting $H_0^{\rm local}$ would be approximately $69.25^{+0.32}_{-0.26}$, $68.89^{+0.21}_{-0.19}$, and $72.31^{+0.27}_{-0.24}$~km/s/Mpc for the exponential, Gaussian, and Maxwell–Boltzmann profiles, respectively. Under these conditions, the exponential void would no longer provide a viable resolution to the Hubble tension, while the Maxwell–Boltzmann profile would predict an $H_0^{\rm local}$ consistent with SH0ES. However, including absolute calibration favours $\beta < 1$ for the Maxwell–Boltzmann void, which shifts its predicted $H_0^{\rm local}$ closer to the \emph{Planck} value.

\subsection{Caveats and future work}

A potential limitation of our approach is the assumption that the void and associated outflow velocities arise within a gravity model featuring faster structure growth than predicted by $\Lambda$CDM (hence faster outflow velocity for given final density). To circumvent the fact that a void with the size and depth of the KBC void cannot arise in $\Lambda$CDM,~\citetalias{Haslbauer_2020} adopted a MOND-based framework. Other gravity models giving faster-than-$\Lambda$CDM structure growth (e.g., theories with new gravitational degrees of freedom or fifth forces) will produce somewhat different outflows for a given density profile. As an alternative strategy, one could adopt a fully $\Lambda$CDM-based void and constrain the rarity of the initial underdensity preferred by the peculiar velocity data, while still assuming the present $\dot{a} = H_0^{\mathrm{Planck}}$ at the background level. It is possible that since our analysis implies a much smaller void than reported by~\citet{Keenan_2013}, such a small void is actually consistent with $\Lambda$CDM, though the measured $H_0^{\mathrm{local}}$ is not.

We note four further caveats. First, we did not vary the initial void depth directly; instead, we introduced a less physical proxy parameter $\beta$ to rescale the predicted velocities and approximate the effects of varying the initial underdensity. This was to remove the computational cost of having to calculate the velocity field in a very large number of models. Second, we assumed that the observer lies along the void's axis of symmetry, a simplification necessitated by the practical constraints of the void model. Third, our inference of the void properties is dependent on the adopted background value of $H_0$. Our choice here, $\dot{a} = H_0^\text{Planck}$, was motivated by the possibility of solving the Hubble tension entirely through void outflows. Adopting instead $\dot{a} \approx 75$~km/s/Mpc would eliminate most if not all the evidence for a local underdensity, as the peculiar velocities would become much smaller. However, it is unclear how such a high present $\dot{a}$ can arise given the CMB power spectrum and the indications from BAO measurements of a faster expansion rate than the \emph{Planck} cosmology at $z \approx 0.5-1$~\citep{DESI_2024, DESI_2025, Mukherjee_2025, Mirpoorian_2025}. Preserving the angular diameter distance to recombination requires a compensatory period with sub-\emph{Planck} expansion rate at $z \la 0.5$. Inferring the local density with less sensitivity to $H_0$ must utilize alternative information such as galaxy number densities~\citep[e.g.,][]{Jasche_2013, Lavaux_2016, Jasche_2019}. Fourth, the velocity structure of any local void will be more complex than the simplified spherically symmetric outflow plus systemic motion model explored here, which may be significant for our field-level results.

Future possibilities also lie in expanding the data used. In this work, we used the \ac{CF4} \ac{TFR} galaxies, 12 of which have absolute distance calibrations from the SH0ES Cepheid calibration~\citep{Riess_2022_comprehensive} after application of our quality cuts. While the \ac{TFR} sample offers full sky coverage, its primary limitation is its shallow depth, with only a small fraction of galaxies at redshifts $\zcmb > 0.05$. Nevertheless, the \ac{TFR} remains attractive because it can be recalibrated straightforwardly, without requiring a full covariance matrix -- as is necessary for datasets like Pantheon+. In contrast, fundamental plane or SNe samples probe to significantly higher redshifts, potentially offering complementary constraints.

If a void were to extend to scales much larger than those currently probed, this would likely result in an unacceptably large bulk flow -- unless the observer were located near the void centre, which becomes increasingly improbable for larger voids. It is statistically implausible for us to reside at the exact centre of a large void: the geometric volume corresponding to such a position is vanishingly small, but observers are expected to be randomly distributed within the cosmic web. For an observer uniformly distributed in volume, the probability density of lying at distance $\RLG$ from the void centre is  $p(\RLG) = 3 \RLG^2 / R_{\rm void}^3$, where $R_{\rm void}$ is the void radius. Thus, for a fixed $\RLG$, the probability falls off as $R_{\rm void}^{-3}$, making the same offset increasingly unlikely in larger voids. Moreover, the dynamics of voids do not single out the centre as a preferential location for haloes, since galaxies inside voids typically experience outflows away from the centre. Thus, although we may lie close to the centre of a void, the hypothesis that we occupy its precise centre is highly unlikely. Even in that case, exponential or Gaussian voids with $\tilde{r}_{\rm void} > 0.5$ would predict $H_0^{\mathrm{local}} \approx 77~\mathrm{km}/\mathrm{s}/\mathrm{Mpc}$, substantially exceeding the SH0ES value of $73.17 \pm 0.86~\mathrm{km}/\mathrm{s}/\mathrm{Mpc}$ and therefore over-resolving the Hubble tension. This problem could be alleviated by scaling down the outflow velocities using $\beta < 1$.

\section{Conclusions}\label{sec:conclusions}

We have investigated the local void model of~\citetalias{Haslbauer_2020} for addressing the Hubble tension, which was motivated by the KBC void~\citep{Keenan_2013}, a reported underdensity in near-infrared galaxy number counts corresponding to an $\approx 20$~per cent matter underdensity extending out to 300~Mpc in radius. While the original analysis by~\citetalias{Haslbauer_2020} did not consider constraints from local peculiar velocity data beyond the \ac{LG}, we introduce and implement this key observational test. Prior to our work,~\citetalias{Mazurenko_2024} calibrated the void model using the measured bulk flow curve of~\citetalias{Watkins_2023}, but their analysis was limited by reliance solely on {these measurements and impacted by an important error, which requires the predictions of~\citetalias{Mazurenko_2024} to be tripled. To address this, we employed the galaxy-by-galaxy flow model of~\citetalias{VF_olympics}, forward-modelling the observed redshift, apparent magnitude, and linewidth of each \ac{TFR} galaxy under the assumption of a predicted peculiar velocity field from the void model. By requiring consistency with the observed redshifts in the \ac{CF4} catalogue, we derived corresponding constraints on the preferred void parameters: its size, systemic velocity relative to the CMB frame, observer offset, and a velocity scaling parameter $\beta$ that effectively captures variations in the initial void depth. Since $\beta$ can be $<0$, our analysis allows the local density to exceed the cosmic average.

Our main conclusions are as follows:
\begin{enumerate}
    \item The \ac{CF4} \ac{TFR} peculiar velocity data favours void sizes that are $7$, $6$, and $15$ per cent of the fiducial size in~\citetalias{Haslbauer_2020}, corresponding to $69$, $61$, and $34~\mathrm{cMpc}$ for the exponential, Gaussian, and Maxwell-Boltzmann profiles, respectively.
    \item Although the void model is flexible enough to reproduce the measured bulk flow curve of~\citetalias{Watkins_2023}, the field-level analysis favours either a flat bulk flow curve for the exponential and Gaussian profiles, or a steeply rising one for the Maxwell-Boltzmann profile. This casts doubt on the bulk flow inference of~\citetalias{Watkins_2023}, which should be simply a summary statistic of the galaxy-by-galaxy distance data and hence produce constraints that are consistent with it. All cases are dominated by a constant external velocity $\bm{V}_{\rm ext}$, so we advise caution in interpreting the Maxwell-Boltzmann bulk flow due to limited data at large distances.
    \item The Gaussian and Maxwell-Boltzmann voids yield ${H_0^{\rm local} \approx 70.4 \pm 0.4~\mathrm{km}/\mathrm{s}/\mathrm{Mpc}}$, which remains $3\sigma$ below the local measurement of ${73.2 \pm 0.9}$~km/s/Mpc~\citep{Breuval_2024}. In contrast, the larger outflow in the exponential profile provides ${H_0^{\rm local} \approx 72.1 \pm 0.8}$~km/s/Mpc, within $1\sigma$ of~\citet{Breuval_2024}.
    \item Without considering any expansion rate or source number count data, Bayesian model selection using the observed velocity field favours the Gaussian and Maxwell-Boltzmann profiles over the exponential profile. The Gaussian profile is only marginally preferred over the Maxwell-Boltzmann profile, despite their drastically different kinematic signatures. The fiducial size void profiles, while disfavoured compared to the smaller void sizes preferred by our analysis, are nonetheless still preferred over a simpler model with constant $\bm{V}_{\rm ext}$ (essentially a void-free control model) in the exponential and Gaussian cases, and only mildly disfavoured in the Maxwell-Boltzmann case.
    \item Given the redshift of the Coma Cluster, the exponential and Gaussian profiles are consistent with its measured distance under the \textit{HST} distance ladder calibration (under-predicting it at $<1\sigma$), while the Maxwell-Boltzmann profile over-predicts it at the $2\sigma$ level.
\end{enumerate}

In summary, we find that allowing the relative void size, external velocity, and velocity scaling to vary yields a better fit across all profiles compared to the best-fit model of~\citetalias{Haslbauer_2020}. This suggests that, although their original model may in principle resolve the Hubble tension, the \ac{CF4} \ac{TFR} data instead favour smaller and possibly deeper void profiles, which for the exponential profile only can still solve the Hubble tension at just over $1\sigma$. For the void to remain as a potentially valid solution to the Hubble tension, it will be important to investigate both the extent to which a smaller void than considered in~\citetalias{Haslbauer_2020} can fit other constraints and the extent to which void models with more complex density profiles can simultaneously fit all observables. This is especially important given that the H0LiCOW analysis suggesting the Hubble tension persists out to quite high redshift is no longer considered valid~\citep{Birrer_2020}, with more modern studies suggesting a more rapid decay to the \emph{Planck} distance--redshift relation~\citep{Jia_2023, Jia_2025}. Lastly, it will also be interesting to develop theoretical frameworks which could create the proposed enhancement to structure on $\ga 100$~Mpc scales relative to $\Lambda$CDM while preserving its success with the CMB anisotropies.

\section*{Data availability}

The \ac{CF4} data is publicly available.\footnote{\url{https://edd.ifa.hawaii.edu/dfirst.php}} The flow model code and all other data will be made available on reasonable request to the corresponding author.

\section*{Acknowledgements}

RS acknowledges financial support from STFC Grant no. ST/X508664/1, the Snell Exhibition of Balliol College, Oxford, and the Center for Computational Astrophysics (CCA) Pre-doctoral Program. HD and IB are supported by a Royal Society University Research Fellowship (grant no. 211046). This project has received funding from the European Research Council (ERC) under the European Union's Horizon 2020 research and innovation programme (grant agreement no. 693024).

We thank Pedro Ferreira, Sebastian von Hausegger, Mike Hudson, Pavel Kroupa, Guilhem Lavaux, Sergij Mazurenko, Jos\'e Antonio N\'ajera, and Tariq Yasin for useful inputs and discussions. We also thank Jonathan Patterson for smoothly running the Glamdring Cluster hosted by the University of Oxford, where the data processing was performed.

\bibliographystyle{mnras}
\bibliography{ref}

\begin{appendix}

\section{Limits from the CMB quadrupole}\label{sec:CMB_quadrupole}

In addition to the bulk flow curve, another important constraint on any solution to the Hubble tension is its ability to retain consistency with the CMB power spectrum while matching the high local $cz'$. A local void solution has an advantage in this respect because it does not alter the early universe or the angular diameter distance to the CMB~\citep[Appendix~A of][]{Banik_2025_BAO}. However, since we cannot be located exactly at the centre of a local void, its potential would gravitationally lens the CMB anisotropically, especially at low multipole moments/large angular scales. The impact on the CMB monopole due to the height of the local potential would be very small~\citepalias[Section~5.3.3 of][]{Haslbauer_2020}. Our peculiar velocity creates a dipole in the CMB, which can be explained by the void in about 2~per cent of its volume (see their Section~4.2.3).

Given this consistency with the CMB dipole, the impact on higher multipoles would be small according to~\citet{Alnes_2006}, whose Equations~$37-39$ ``imply that it is impossible to obtain sufficiently large values for the quadrupole and octopole [to significantly affect the observations] as long as the dipole is within the limits set by the data.'' They show that the impact on the CMB multipole moment $\ell$ is $\mathcal{O} \left( \alpha^\ell \right)$, where
\begin{eqnarray}
    \alpha = \frac{d \left( h_{\mathrm{in}} - h_{\mathrm{out}} \right)}{3000 \, \text{Mpc}} \, ,
\end{eqnarray}
$h$ indicates a value in units of $100 \, \mathrm{km} / \mathrm{s} / \mathrm{Mpc}$, $h_{\mathrm{out}} = \dot{a}$, $h_{\mathrm{in}} = cz'$ within the void, and the constant of 3000~Mpc denotes the Hubble distance $d_H \equiv c/H$ for the case $h = 1$. Any local void solution to the Hubble tension must have $h_{\mathrm{in}} - h_{\mathrm{out}} \approx 0.07$~\citep{Valentino_2025}. Since~\citetalias{Mazurenko_2024} found that we need to be $\approx 150$~Mpc from the void centre to match the bulk flow curve, the CMB multipole moments approximately become powers in the parameter $\alpha = 3.5 \times 10^{-3}$, with the quadrupole being $\mathcal{O} \left( \alpha^2 \right) \approx 10^{-5}$ and the octopole being $\mathcal{O} \left( \alpha^3 \right) \approx 10^{-7}$. It is clear that even if we were to consider substantial changes to our vantage point to account for their mistake with how bulk flows are defined (Section~\ref{sec:bf_model}), the octopole contribution would not be discernible given the extra factor of $\sqrt{16 \mathrm{\pi}/175} = 0.54$~\citep[Equation~39 of][]{Alnes_2006}, making the void contribution much less than the $\mathcal{O} \left( 10^{-5} \right)$ intrinsic octopole in the CMB. Its higher-order multipoles would be even less affected by a local void. However, the quadrupole deserves more attention given the comparable magnitudes.

The quadrupole contribution given by Equation~38 of~\citet{Alnes_2006} is as follows:
\begin{eqnarray}
    a_{20} = \sqrt{\frac{16 \mathrm{\pi}}{45}} \alpha^2.
\end{eqnarray}
Since we will vary the void parameters and our vantage point, we estimate $\alpha$ and thus $a_{20}$ for each combination of parameters. We assume that for a model to be viable,
\begin{eqnarray}
    a_{20} < 10^{-5}.
    \label{eq:CMB_quadrupole_limit}
\end{eqnarray}
This ensures that the contribution from a local void is smaller than the intrinsic power in the CMB quadrupole. The observed CMB quadrupole must be a combination of the intrinsic quadrupole and \emph{anisotropic} lensing by the potential of any local void the late-time observer might be located within, so too large a contribution from the latter would cause difficulties matching the observed CMB anisotropies. However, even the very extreme void shown in Figure~2 of~\citet{Nistane_2019} would only marginally be detectable with \textit{Planck}, highlighting that this argument can only be used to rule out a significant offset from a substantial underdensity. We will see later that despite this limitation, Equation~\ref{eq:CMB_quadrupole_limit} places an important constraint on the allowed solutions. These constraints are significant for the fiducial void size, requiring the observer to be $\la 90$~Mpc from the void centre in all three profiles. We will use this to rule out some solutions in~\cref{sec:results_fiducial_bf}. On the other hand, the CMB quadrupole constraints become irrelevant for the smaller void sizes preferred by our main field-level analysis, as such voids would not produce any detectable imprint on the CMB for observer locations that plausibly match the local velocity field.

\section{Fiducial size results}\label{sec:results_fiducial}

\subsection{Bulk flow constraints}\label{sec:results_fiducial_bf}

While the main body of this paper focuses on results obtained allowing the void size to vary, here we revisit the fiducial-size models by fixing the size to the values used in~\citetalias{Haslbauer_2020} and varying only the vantage point and external velocity, whose direction is set parallel to the bulk flow of~\citetalias{Watkins_2023}. The results of this fixed-size analysis are presented in~\cref{fig:fiducial_void_chisq_combined}, where we show that acceptable fits to the~\citetalias{Watkins_2023} bulk flow curve can still be obtained, but only within a narrow range of configurations.

\begin{figure*}
    \centering
    \includegraphics[width=\textwidth]{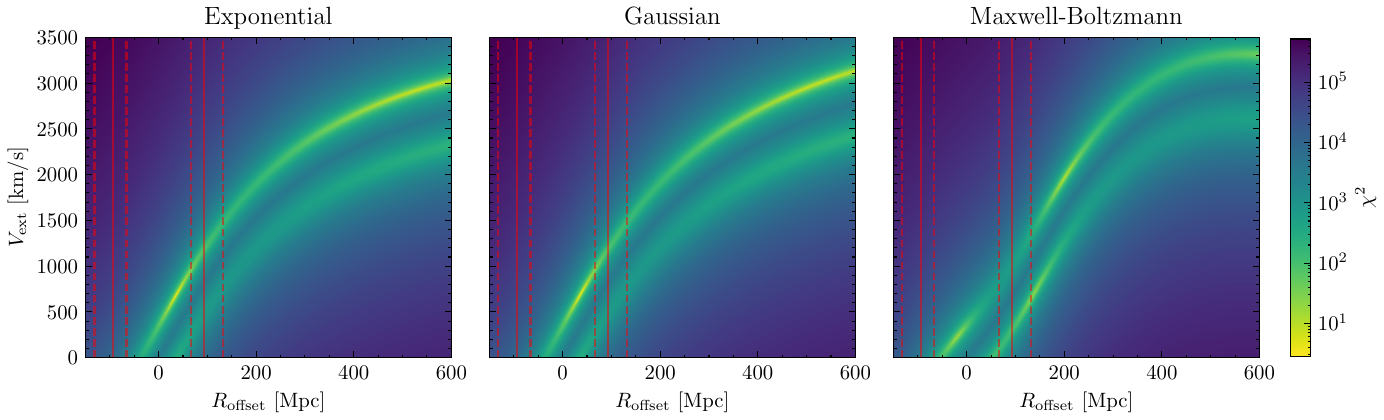}
    \caption{$\chi^2$ as a function of observer offset from the centre of the void ($\RLG$) and external velocity magnitude ($V_{\rm ext}$) for exponential, Gaussian, and Maxwell-Boltzmann void profiles (left, middle, and right panel, respectively). The solid vertical lines mark CMB quadrupole consistency limits, while the dashed lines vary the limit on $a_{20}$ by a factor of 2 from the nominal $10^{-5}$ (see Appendix~\ref{sec:CMB_quadrupole}). In all three void profiles, the outer solution is excluded, but the inner solutions remain viable.}
    \label{fig:fiducial_void_chisq_combined}
\end{figure*}

In the left and middle panels of~\cref{fig:fiducial_void_chisq_combined}, we show the $\chi^2$ landscape for the exponential and Gaussian profile, respectively. Both profiles give quite similar results. In each case, there are two broad valleys in $\chi^2$. These arise because the bulk flows reported by~\citetalias{Watkins_2023} are actually quite \emph{small} in comparison to the typical peculiar velocity in the void models~\citepalias[see Appendix~A of][]{Mazurenko_2024}. Our vantage point needs to be some distance from the void centre so that the void outflow velocity $V_{\rm{outflow}}$ cancels $V_{\rm ext}$ given these are in opposite directions~\citepalias[in their figures, the void moves to the left and the vantage point is on the right; see also Figure~8 of][]{Haslbauer_2020}. For a given $V_{\rm ext}$, perfect cancellation occurs at some $R_{\rm offset}$. However, we do not want a perfect cancellation because $V_b \neq 0$. Since the orientation of the local void model is not known independently of the observed $\bm{V}_b$, a good fit to the observed $V_b$ on the smallest probed scales arises when $V_{\rm{outflow}} - V_{\rm{void}} = \pm V_b$, though the different signs require opposite relative orientations between $\bm{V}_b$ and the direction towards the void centre. The $+$~case corresponds to a larger $R_{\rm offset}$ because we expect $V_{\rm{outflow}}$ to increase with $R_{\rm offset}$ by about 7~km/s/Mpc in order to solve the Hubble tension. In both panels, this valley in $\chi^2$ does not contain a viable solution because $\chi^2 \gg 100$. Focusing on the smaller $R_{\rm offset}$ (or upper) valley, we see that there are two viable (bright yellow) regions in the exponential and Gaussian panels of~\cref{fig:fiducial_void_chisq_combined}. In principle, both represent viable fits to the bulk flow curve of~\citetalias{Watkins_2023}. However, the solution at large $R_{\rm offset}$ places us much too far from the void centre to be consistent with the observed CMB quadrupole. In the following, we therefore only consider the region towards the lower left in voids with the fiducial size from~\citepalias{Haslbauer_2020}.

In the right panel of~\cref{fig:fiducial_void_chisq_combined}, we present the $\chi^2$ landscape for the Maxwell-Boltzmann profile calibrated against the bulk flow curve of~\citetalias{Watkins_2023}. Given that there are nine data points and two degrees of freedom, a viable model should ideally achieve $\chi^2 \approx 7$, corresponding to the bright yellow contours. We identify three regions that provide a good fit. The two solutions at $\RLG \approx 200$~Mpc imply a substantially off-centre observer, where the CMB quadrupole would experience a significant contribution due to anisotropic lensing by the void (Appendix~\ref{sec:CMB_quadrupole}). The solid vertical lines indicate the offset at which this contribution reaches $10^{-5}$, a rough estimate of the intrinsic quadrupole in the CMB (Eq.~\ref{eq:CMB_quadrupole_limit}). The dashed (dot-dashed) vertical lines represent where the contribution is half (twice) this value. Even allowing for some flexibility in applying the CMB quadrupole constraint, it is clear that both ``outer'' solutions are problematic given the excellent agreement of the CMB power spectrum with $\Lambda$CDM~\citep{Planck_2020, Tristram_2024}. Furthermore, such a large offset would inevitably introduce significant anisotropy in galaxy number counts. Given these constraints, we focus only on the ``inner'' solution at $\RLG \approx -20$~Mpc, where the lensing contribution to the CMB quadrupole would be very small.

\begin{figure}
    \centering
    \includegraphics[width=\columnwidth]{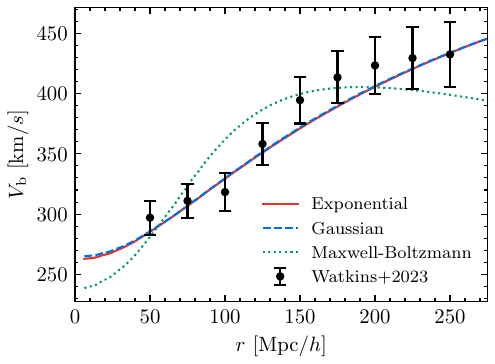}
    \caption{The bulk flow curve for parameters corresponding to the local minima in the $\chi^2$ landscape with acceptable $\chi^2$ for the exponential (red), Gaussian (blue), and Maxwell-Boltzmann (green) void density profile. The plotted curves correspond to the lower $\RLG$ (``inner'') solutions summarized in~\cref{tab:chi_sq_values_fid_size}.}
    \label{fig:Bulkflows_best_fits}
\end{figure}

To demonstrate that the low $\chi^2$ solutions are indeed good fits to the observed bulk flow curve, we show the predicted bulk flow curve in each case (\cref{fig:Bulkflows_best_fits}), with the corresponding parameters summarized in~\cref{tab:chi_sq_values_fid_size}. Only the inner solution for each profile can be considered viable given the CMB quadrupole constraint (Appendix~\ref{sec:CMB_quadrupole}), so we only plot the inner solutions in~\cref{fig:Bulkflows_best_fits}. It is interesting to note that the fiducial void size yields a better fit to the bulk flow curve in terms of $\chi^2$ than the smaller void sizes studied earlier in~\cref{sec:res_bulk_flow_size}, where the minimum was $\chi^2 = 9.1$, compared to $\chi^2 = 2.9$ with the fiducial size.

\begin{table*}
    \centering
    \begin{tabular}{ccccc}
    \hline
    &  \multicolumn{4}{c}{Density profile} \\
    Quantity & Vantage point & Exponential & Gaussian & Maxwell-Boltzmann \\ \hline
    \multirow{2}{*}{$R_{\rm offset}$ (Mpc)} & Inner & 36 & 37 & -19 \\
    & Outer & 538 & 530 & 200 \\
    \multirow{2}{*}{$V_{\mathrm{ext}}$ (km/s)} & Inner & 689 & 699 & 234 \\
    & Outer & 2940 & 3012 & 1966\\
    \multirow{2}{*}{$\chi^2$ (tension)} & Inner & 2.9 ($0.13\sigma$) & 2.9 ($0.13\sigma$) & 18.7 ($2.61\sigma$) \\
    & Outer & 8.6 ($1.07\sigma$) & 8.6 ($1.07\sigma$)& 11.4 ($1.55\sigma$) \\ \hline
    \hline
    \end{tabular}
    \caption{Parameters corresponding to the local minima in the $\chi^2$ landscape with acceptable $\chi^2$ for each density profile. The $\chi^2$ values have been converted to an equivalent tension for a single Gaussian variable assuming 7 degrees of freedom, since there are 9 data points and we used 2 free parameters to optimize the fit. The outer solutions are rejected on the basis of the CMB quadrupole constraint (Appendix~\ref{sec:CMB_quadrupole}), so only the inner solutions shown here are plotted in~\cref{fig:Bulkflows_best_fits}.}
    \label{tab:chi_sq_values_fid_size}
\end{table*}

\subsection{Galaxy-by-galaxy constraints}

We now fix the void size to $\tilde{r}_{\rm void} = 1$ and repeat the field-level analysis of the main part of the paper.
For a more straightforward comparison to~\citetalias{Haslbauer_2020}, we also fix $\beta = 1$. However, we note that these are strong assumptions -- in~\cref{sec:results_varying}, we have already shown that a value of 1 is not preferred for either parameter. Unlike the bulk flow constraint in Appendix~\ref{sec:results_fiducial_bf}, we also treat the direction of $\bm{V}_{\rm ext}$ in Galactic coordinates $\left(\ell_{\rm ext},\,b_{\rm ext} \right)$ as a free parameter.

The inferred values of the void model parameters are shown in~\cref{fig:fiducial_void}. We infer a direction for $\bm{V}_{\rm ext}$ that is well-aligned with the bulk flow direction reported in Table~1 of~\citetalias{Watkins_2023} for both the exponential and Gaussian profile, but not for the Maxwell-Boltzmann profile. The inferred $R_{\rm offset}$ and $V_{\rm ext}$ are highly correlated, which is to be expected given the narrow valleys in the $\chi^2$ landscape evident in~\cref{fig:fiducial_void_chisq_combined}. As discussed in~\cref{sec:results_fiducial_bf}, these valleys arise because of the need for outflow from the void to almost cancel $\bm{V}_{\rm ext}$ and give a relatively low total $\bm{V}_{\rm pec}$ at our location, implying a degeneracy between $\bm{V}_{\rm ext}$ and $\RLG$, which sets the outflow velocity.

\begin{figure}
    \centering
    \includegraphics[width=\columnwidth]{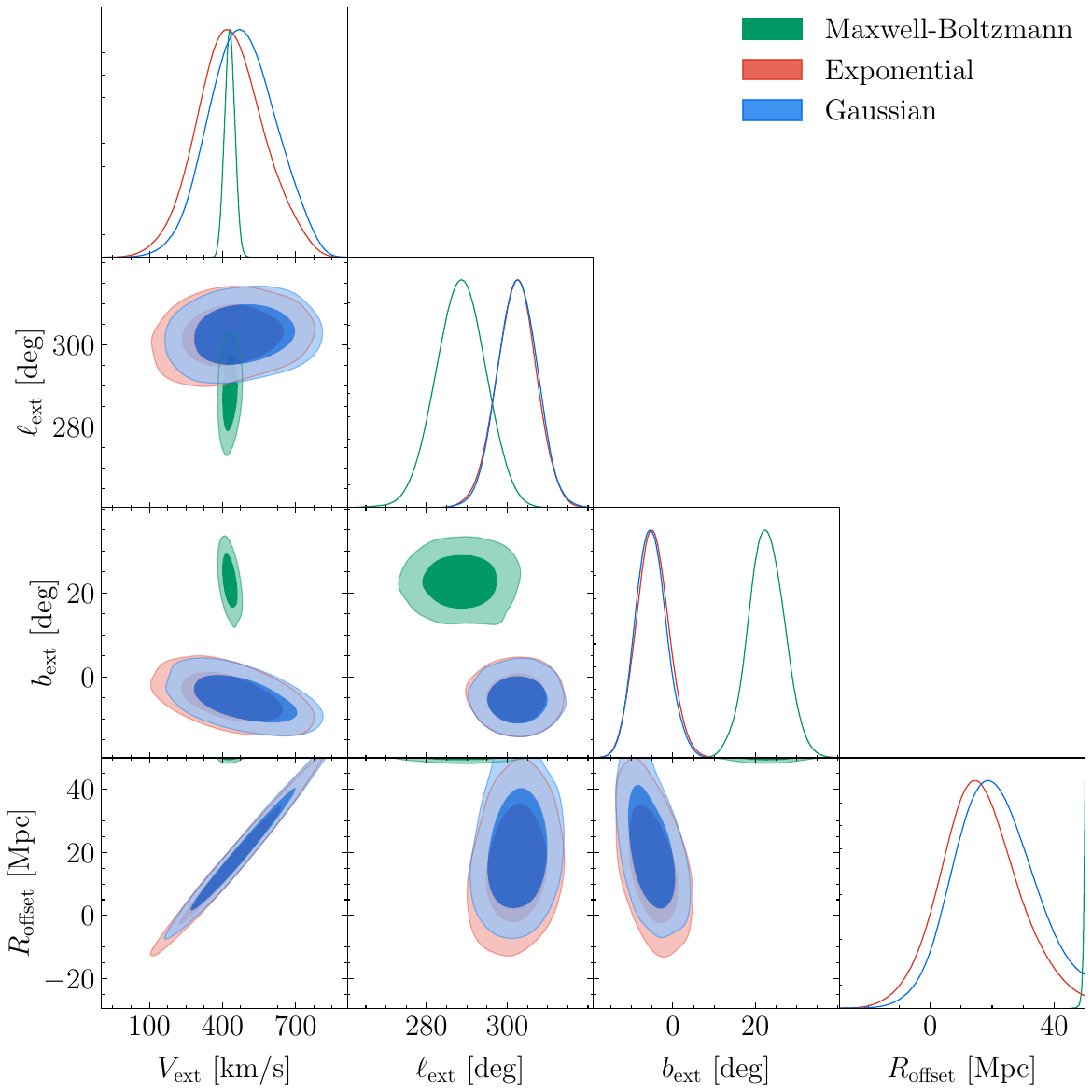}
    \caption{The inferred void model parameters obtained by fitting the flow model described in~\cref{sec:flow_model} to the \ac{CF4} data using the \emph{fiducial void size}~\citepalias{Haslbauer_2020}, assuming either an exponential (red), Gaussian (blue), or Maxwell-Boltzmann (green) void profile. An important parameter that we sample is the external velocity, which lies in the direction $\left( \ell_{\rm ext}, \, b_{\rm ext} \right)$ in Galactic coordinates and has magnitude $V_{\rm ext}$. The other parameter we sample is $\RLG$, the distance of our vantage point from the void centre in the direction opposite to $\bm{V}_{\rm ext}$.}
    \label{fig:fiducial_void}
\end{figure}

However, \cref{fig:void_fiducial_comparison} shows that the best-fit locations of $V_{\rm ext}$ and $\RLG$ differ from the ones inferred by calibrating the void model to the bulk flow curve of~\citetalias{Watkins_2023}. In particular, we infer a smaller $\RLG$ and correspondingly smaller $V_{\rm ext}$. Even so, the tension between the field-level and bulk-flow-only constraint is $1.7\sigma$ ($1.6\sigma$) for the exponential (Gaussian) profile, suggesting no strong disagreement. As with the varying void size analysis, it is surprising that the uncertainties from the bulk flow constraint are significantly tighter than those from the field-level inference, despite the latter having in principle access to more information than the single summary statistic of the bulk flow. This likely reflects either an underestimation of the bulk flow uncertainties~\citep{Whitford_2023} or the neglect of correlations between the bulk flow measurements when fitting the void model. In contrast, the field-level analysis for the Maxwell-Boltzmann profile indicates $\RLG > 50$~Mpc and $V_{\rm ext} > 400$~km/s, in tension with the bulk flow constraints as these favour $\RLG \approx -19$~Mpc and $V_{\rm ext} \approx 234$~km/s.

Upon comparing the Bayesian evidence across the three fiducial void profiles, we find that the Maxwell-Boltzmann model is strongly disfavoured. The exponential and Gaussian profiles yield $\log \mathcal{B}$ values of 83.5 and 83.0, respectively, relative to the Maxwell-Boltzmann case. Given this decisive evidence, the fiducial Maxwell-Boltzmann profile is ruled out by the \ac{CF4} data and is not considered further in this analysis. These results are consistent with the bulk flow-only analysis, which cannot distinguish between the exponential and Gaussian profiles, but yields a $\chi^2$ value worse by 15.8 for the Maxwell-Boltzmann profile~\citepalias[see also][]{Mazurenko_2024}.

\begin{figure}
    \centering
    \includegraphics[width=\columnwidth]{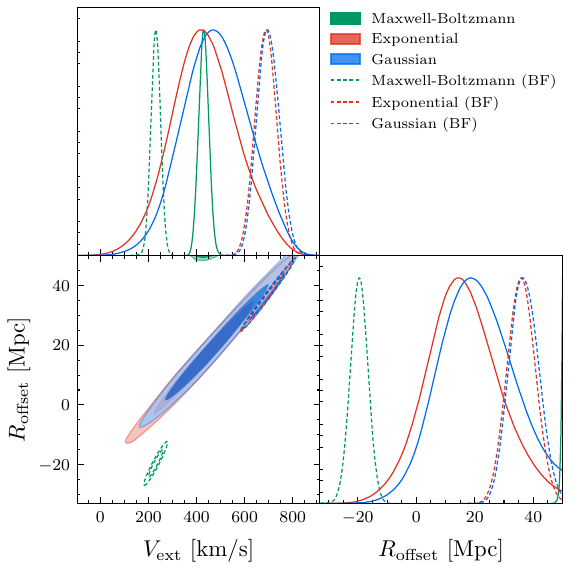}
    \caption{Comparison of $V_{\rm ext}$ and $\RLG$ between the field-level (solid) and bulk flow (dashed) constraints for the fiducial void size. The exponential and Gaussian profiles are in good agreement, unlike the Maxwell-Boltzmann profile. It is surprising that the bulk flow curve based on just nine data points gives a much tighter inference.}
    \label{fig:void_fiducial_comparison}
\end{figure}

Having inferred $\bm{V}_{\rm ext}$ and $\RLG$, we now show the void density profile, outflow velocity curve, and bulk flow curve in \cref{fig:fiducial_void_stats}. Its left panel shows the expected 20~per cent underdensity at the void centre for the exponential and Gaussian profiles. The density rises towards the cosmic mean further out, but only gradually, since the fiducial void size is 1030~cMpc. On the other hand, the Maxwell-Boltzmann profile exhibits a growing underdensity out to ${\approx 200~\mathrm{Mpc}/h}$, after which the void gradually becomes shallower and approaches the cosmic mean density. The middle panel shows that the outflow velocity at first rises with a rate of ${\approx 10}$~km/s/Mpc, before gradually becoming less steep as the density gets closer to the cosmic mean. At the void centre, we expect the outflow velocity to rise somewhat faster than the magnitude of the Hubble tension (${cz' - \dot{a}}$) because the observed $cz'$ is an average over a large range of radii, reducing the average value compared to that at the void centre. The right panel shows that our analysis prefers exponential and Gaussian models where the bulk flow curve rises less steeply than reported by~\citetalias{Watkins_2023}. This is related to the fact that our field-level analysis places us closer to the void centre, where the bulk flow curve would be nearly flat. The dotted curves, which show the void-only outflow, highlight that the bulk flow in these models is largely due to $\bm{V}_{\rm ext}$, since removing it leads to only a very small $V_b$. However, $\bm{V}_b$ does not entirely arise from $\bm{V}_{\rm ext}$ either---if it did, we would get $\bm{V}_b = \bm{V}_{\rm ext}$ at all radii, leading to a flat bulk flow curve. The dashed lines provide a check of our earlier result that the predicted bulk flow curve in any of the considered void profiles can provide a good match to that reported by~\citetalias{Watkins_2023}, if the void model is calibrated to it. The bulk flow curves inferred by our field-level analysis provide a reasonably good match to that reported by~\citetalias{Watkins_2023} for the exponential and Gaussian profiles, though the~\citetalias{Watkins_2023} result rises a bit more steeply. On the other hand, in the case of the Maxwell-Boltzmann profile, $\bm{V}_{\rm ext}$ cancels the void outflow at large radii according to our field-level analysis, leading to a negligible bulk flow. This is strongly incompatible with~\citetalias{Watkins_2023}, and indeed the fiducial Maxwell-Boltzmann void is strongly rejected by our model comparison metric.

\begin{figure*}
    \centering
    \includegraphics[width=\textwidth]{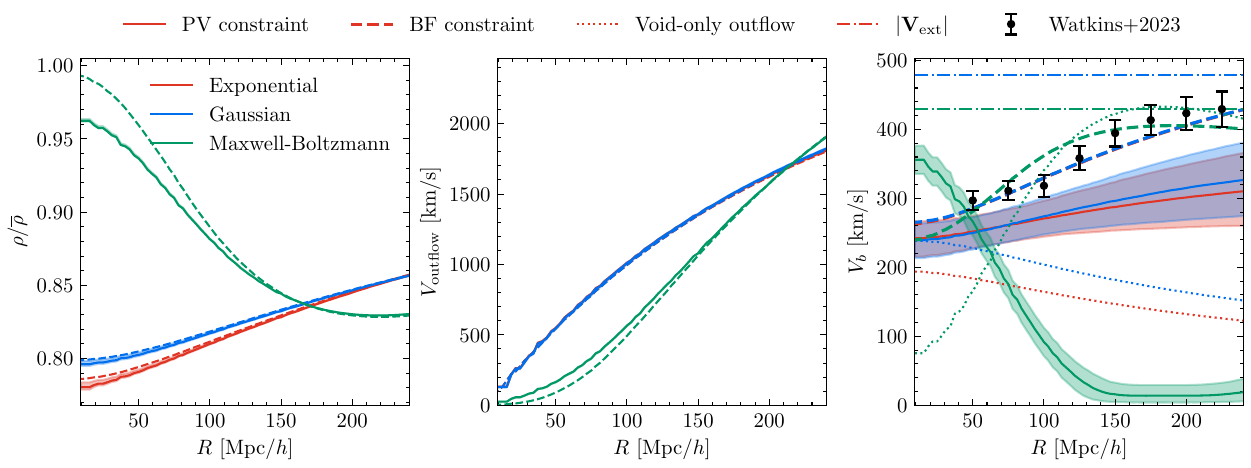}
    \caption{The inferred density field of the \emph{fiducial size} void (left panel), the outflow velocity curve from the void centre (middle panel), and the bulk flow curve (right panel), using the void parameters and their uncertainties as inferred by our flow analysis. We show results for the exponential (Gaussian; Maxwell-Boltzmann) profile using red (blue; green) shaded bands, which indicate the $1\sigma$ uncertainty. In the right panel, the dot-dashed lines show $V_{\rm ext}$ (the exponential and Gaussian profile curves overlap) and the dotted lines show the bulk flow without considering $\bm{V}_{\rm ext}$, illustrating its impact. The dashed lines show the bulk flow curve in the model which best fits that reported by~\citetalias{Watkins_2023}, which we show as black points with uncertainties.}
    \label{fig:fiducial_void_stats}
\end{figure*}

\end{appendix}

\bsp
\label{lastpage}
\end{document}